\documentclass[reprint,amsmath,amssymb,prl,aps,noeprint,floatfix,nolinenumbers,superscriptaddress,longbibliography]{revtex4-2}

\usepackage[english]{babel}
\usepackage[utf8]{inputenc}
\usepackage[colorinlistoftodos, color=green!40, prependcaption]{todonotes}
\usepackage{float}
\usepackage{graphicx}
\usepackage{yhmath}
\usepackage{subfigure}
\usepackage{mathdots}
\usepackage{MnSymbol}
\usepackage{dsfont}
\usepackage{soul}
\usepackage[normalem]{ulem}
\usepackage{hyperref}

\hypersetup{
    unicode=true, 
    plainpages=false,
    colorlinks=true,
    linkcolor=blue,
    citecolor=blue,
    filecolor=blue,
    urlcolor=blue
}

\usepackage{mathtools}

\DeclarePairedDelimiterX\braket[2]{\langle}{\rangle}{#1\,\delimsize\vert\,\mathopen{}#2}

\newcommand{\md}{\mathrm{d}}

\newcommand{\sgn}{\mathrm{sgn}}

\begin{document}
\title{Vector Ising Spin Annealer for Minimizing  Ising Hamiltonians}
\author{James S. Cummins${}^1$ and Natalia G. Berloff}
\email[correspondence address: ]{N.G.Berloff@damtp.cam.ac.uk}
\affiliation{Department of Applied Mathematics and Theoretical Physics, University of Cambridge, Wilberforce Road, Cambridge CB3 0WA, United Kingdom}

\date{\today}

\begin{abstract}
We introduce the Vector Ising Spin Annealer (VISA), a framework in gain-based computing that harnesses light-matter interactions to solve complex optimization problems encoded in spin Hamiltonians. Traditional driven-dissipative systems often select excited states due to limitations in spin movement. VISA transcends these constraints by enabling spins to operate in a three-dimensional space, offering a robust solution to minimize Ising Hamiltonians effectively. Our comparative analysis reveals VISA's superior performance over conventional single-dimension spin optimizers, demonstrating its ability to bridge substantial energy barriers in complex landscapes. Through detailed studies on cyclic and random graphs, we show VISA's proficiency in dynamically evolving the energy landscape with time-dependent gain and penalty annealing, illustrating its potential to redefine optimization in physical systems.

\end{abstract}

\maketitle

\section{Introduction}
In pursuing advancements in digital computing, mainly aimed at addressing the complexities inherent in AI and large-scale optimization problems, the inherent limitations of the traditional von Neumann architecture come to the forefront. These limitations are characterized by the escalating costs required for incremental performance improvements, prompting a pivotal shift toward more sustainable computational paradigms \cite{thompson2021decline}. In this evolving landscape, two primary paths have emerged: refining algorithms within the existing computational frameworks and exploring novel hardware paradigms grounded in physics principles, with a notable focus on exploiting light-matter interactions.

Algorithmic enhancements, while valuable, offer incremental improvements. In contrast, the exploration of alternative hardware paradigms holds the promise of a substantial shift, leveraging the fundamental principles of physics—such as minimizing entropy, energy, and dissipation \cite{vadlamani2020physics}—and integrating quantum phenomena like superposition and entanglement \cite{farhi}. This innovative approach aims to transcend the conventional computing model's limitations, tapping into the intrinsic computational potential of physical systems to address complex optimization challenges currently beyond traditional methods' reach.

Central to this innovative trajectory is the integration of analogue, physics-based algorithms and hardware, which involve  translating complex optimization problems into universal spin Hamiltonians, like the classical Ising and XY models \cite{lucas2014ising, berloff2017realizing, kalinin2020polaritonic}. This translation process involves embedding the structure of the problem into the spin Hamiltonian's coupling strengths and targeting the ground state as the solution, with the physical system tasked with discovering this state. The efficiency and accuracy of this mapping process are vital, as they ensure the scalability of computational efforts, enabling these problems to remain manageable despite increasing complexity \cite{Cubitt_universality}.

Gain-based computing (GBC) using coupled light-matter emerges as a novel disruptive computing platform distinct from gate-based and quantum or classical annealing approaches. GBC combines optical light manipulations with advancements in laser technology and spatial light modulators, facilitating parallel processing across multiple channels with significant nonlinearities and high-energy efficiency. The operational principle of GBC—increasing pumping power (annealing), followed by symmetry breaking and gradient descent—relies on wave coherence and synchronization to lead the system to a state that minimizes losses naturally. Recent developments in physics-based hardware exploiting GBC principles have showcased diverse technologies. These range from optical parametric oscillator based coherent Ising machines (CIMs) \cite{mcmahon2016fully, inagaki2016coherent, yamamoto2017coherent, honjo2021100}, memristors \cite{cai2020power}, and laser systems \cite{babaeian2019single, pal2020rapid, parto2020realizing}, to photonic simulators \cite{pierangeli2019large, roques2020heuristic}, polaritons \cite{berloff2017realizing, kalinin2020polaritonic}, photon condensates \cite{vretenar2021controllable}, and novel computational architectures like Microsoft's analogue iterative machine \cite{kalinin2023analog} and Toshiba's simulated bifurcation machine \cite{goto2021high}. These platforms have been instrumental in minimizing programmable spin Hamiltonians, demonstrating efficacy across a spectrum of complex optimization challenges, including machine learning \cite{date2021qubo}, financial analytics \cite{gilli2019numerical}, and biophysical modelling \cite{pierce2002protein, dill2008protein}.

The Ising Hamiltonian, originating from statistical physics, is a natural model that can be implemented in these platforms. It describes interactions between spins in a lattice, providing a fundamental model for understanding complex systems and can be written as
\begin{equation}
H_{\rm I} = - \sum_{ij} J_{ij} s_i s_j - \sum_{i} h_i s_i,
\end{equation}
where $s_{i} = \pm 1$ represents the spin, and $h_i$ is the external magnetic field. $J_{ij}$ is the interaction strength between the $i$-th and $j$-th spins, and is encoded in the coupling matrix $\mathbf{J}$. The Ising model's exploration of ground states in physical systems parallels the search for minimum cost functions in optimization. At the same time, its energy landscapes mirror the loss landscapes in machine learning, offering a unique perspective on optimization problems.

At the core of the hardware representation of Ising Hamiltonians lies the utilization of scalar soft-spins within soft-spin Ising models, which effectively change energy barriers inherent in classical hard-spin Ising Hamiltonians, $H_{\rm I}$. The transition from the hard-spin Ising Hamiltonians $H_{\rm I}$ to the soft-spin Ising models consists of considering binary spins $s_i$ as signs of continuous variables $x_i$ - amplitudes - that bifurcate from vacuum state ${\bf x} = 0$ guided by the gain increase. While facilitating enhanced problem-solving capabilities, this approach occasionally encounters the obstacle of trajectory trapping within local minima due to the escalating energy barriers as gain increases.

Therefore, the challenge of navigating energy landscapes with numerous local minima, especially during amplitude bifurcation, necessitates an innovative approach to avoid the system being trapped in local minima. In this paper, we introduce the Vector Ising Spin Annealer (VISA), a model that integrates the benefits of multidimensional spin systems and soft-spin gain-based evolution. Unlike traditional single-dimension semi-classical spin models, VISA utilizes three soft modes to represent vector components of an Ising spin, allowing for movement in three dimensions and offering a robust framework for accurately determining ground states in complex optimization problems. Our comparative studies underscore the enhanced performance of VISA compared to established models with scalar spins, such as Hopfield-Tank networks, coherent Ising machines, and the spin-vector Langevin model (SVL). VISA showcases a notable ability to overcome significant energy barriers and effectively navigate through intricate energy landscapes. We illustrate VISA's effectiveness by tackling Ising Hamiltonian problems across various graph structures. Our examples include analytically solvable 3-regular circulant graphs, more complex circulant graphs, and random graphs where minimizing the Ising Hamiltonian is known to be an NP-hard problem. In our discussions, when we refer to ``solving a graph A'' or ``minimizing a graph A'', we use shorthand for the more detailed process of ``finding the global minimum of the Ising Hamiltonian on graph A''. This terminology simplifies our reference to determining the lowest possible energy state for the Ising model applied to a specific graph structure.

\section{Vector Ising Spin Annealer}
The VISA model is a semi-classical, three-dimensional soft-spin Ising model. It employs annealing of the loss landscape, symmetry-breaking, bifurcation, gradient descent, and mode selection to drive the system toward the global minimum of the Ising Hamiltonian. Here, we represent Ising soft-spins as continuous vectors in three-dimensional space ${\bf x}_i = (x_{1i}, x_{2i}, x_{3i})$ that are free to move in that space. VISA may be physically realized using amplitudes of a network of coupled optical oscillators. In these optical-based Ising machines, vectors of Ising spins are represented by three soft-spin amplitudes. In total, $3N$ amplitudes are required to minimize an $N$-spin Ising Hamiltonian. This approach is somewhat complimentary to the recently proposed ``dimensionality annealing'' where soft Ising amplitudes are considered as coordinates of the multidimensional spins such as XY or Heisenberg spins \cite{calvanese2022multidimensional, strinati2024hyperscaling}. While the main idea is to exploit the advantages of the higher dimensionality, VISA aims at the Ising minimization rather than hyper-dimensional spin systems \cite{calvanese2022multidimensional, strinati2024hyperscaling}.

To formulate the Hamiltonian that is capable of representing the dynamics of the GBC, we incorporate the term that is convex and dominates when the effective gain (loss) is large and negative, the Ising term that is minimized when the gain is large and positive, and a term that aligns the spins at the end of the process so that their direction can be associated with the binary spin $s_i = \pm 1$. We write, therefore, the VISA Hamiltonian as the sum of three terms $H_{\rm VISA} = H_1 + H_2 + H_3$, where
\begin{align}
    H_1 & = \frac{\alpha}{4} \sum_{i=1}^N (\gamma_i (t) - |{\bf x}_i|^2)^2, \label{H1} \\
    H_2 & = - \frac{1}{2} \sum_{i, j = 1}^N J_{ij} {\bf x}_i \cdot {\bf x}_j, \label{H2} \\
    H_3 & = \frac{P(t)}{2} \sum_{i, j = 1}^N |{\bf x}_i \times {\bf x}_j|^2, \label{H3}
\end{align}
with hyperparameter $\alpha$. As the effective gain $\gamma_i (t)$ increases with time $t$ from negative (effective losses) to positive (effective gain) values, $H_1$ anneals between a convex function with the minimum at ${\bf x}_i = {\bf 0}$ to nonzero amplitudes. $H_2$ coincides with the Ising Hamiltonian when all ${\bf x}_i$ have unit magnitude when projected along the same direction. Finally, $H_3$ is a penalty term with time-dependent magnitude $P(t)$ to enforce collinearity between ${\bf x}_i$ and ${\bf x}_j$ at the end of the gain-induced landscape change. At which time, the condition on the amplitudes $|{\bf x}_i| = 1$ is imposed by the feedback on the gain realised by $\dot{\gamma}_i = \varepsilon (1 - |{\bf x}_i|^2) $ \cite{kalinin2018networks}. Analogously to CIM operation, as $\gamma_i (t)$ grows from negative to positive values, $H_{\rm VISA}$ anneals from the dominant convex term $H_1$ that is minimized at ${\bf x}_i = (0, 0, 0)$ for all $i$, to the minimum of $H_2 + H_3$ with $|{\bf x}_i| = 1$ via symmetry-breaking bifurcation. Concurrently, as $P(t)$ increases from $P(0) = 0$ to sufficiently large $P(T) > 0$, the soft-spins vectors become collinear. The Ising spins are calculated by projecting the three-dimensional vectors along the axis ${\bf k}$ they have centred around and taking signs of the resultant scalar $s_i = \sgn ( {\bf x}_i \cdot {\bf k} )$. At $t = T$, the target hard-spin Ising Hamiltonian $H_{\rm I}$ is minimized.

The Hamiltonian $H_{\rm VISA}$ is $4$-local due to the $H_3$ term. Some optical hardware can directly encode such interactions \cite{stroev2021discrete}. At the same time, it is possible to reduce the $4$-local Hamiltonian to quadratic Hamiltonian by substituting the product of two variables with a new one while adding the appropriate penalty terms. For instance, the quartic interaction term $x_i x_j x_k x_m$ can be reduced to a triplet by introducing a new variable $y$ instead of $x_k x_m$ and the term $q_{km}(y) = Q(x_k x_m - 2 x_k y - 2x_m y + 3 y)$ (known as Rosenberg polynomial \cite{rosenberg1975reduction}) in the Hamiltonian where $Q > 0$ is the penalty constant. Note that $q_{km} = 0$ if and only if $x_k x_m = y$ and is strictly positive otherwise. The reduction of the triplet $x_i x_j y$ is done similarly by replacing $x_i x_j$ with a new spin variable $z$ and introducing the penalty term $q_{ij}(z)$ in the Hamiltonian \cite{boros2002pseudo}. 

The governing equations $\Dot{{\bf x}}_i = - \nabla_i H_{\rm VISA}$ represent the gradient descent combined with the temporal change of annealing parameters $\gamma_i(t)$ and $P(t)$ as
\begin{equation} \label{DA_Evo_Eq}
    \begin{split}
     {\bf \dot{x}}_i & = \alpha {\bf x}_i ( \gamma_i (t) - |{\bf x}_i|^2 ) + \sum_{j = 1}^{N} J_{ij} {\bf x}_j \\
    & t) \Big( {\bf x}_i \sum_{j = 1}^{N} |{\bf x}_j|^2 - \sum_{j = 1}^{N} {\bf x}_j ( {\bf x}_i \cdot {\bf x}_j ) \Big),
    \end{split}
\end{equation}
where in the last two terms, the vector triple product has been invoked. The operation of VISA, therefore, relies on the gradient descent of a gain-driven losses landscape. 

\begin{figure}[ht]
\centering
     \includegraphics[width=0.8\columnwidth]{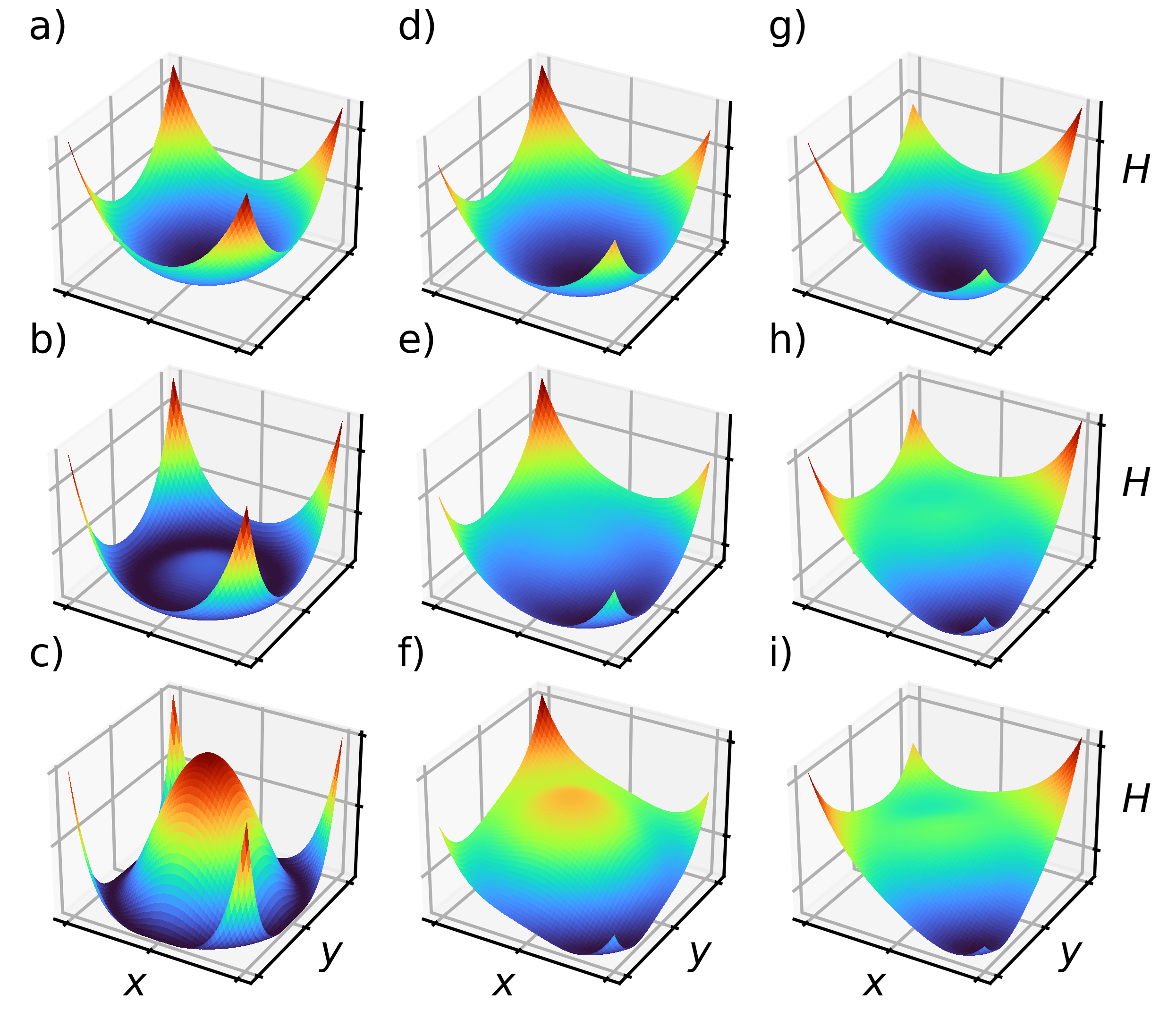}
     \caption{Energy landscape of components of the two-dimensional VISA Hamiltonian for $N = 2$ with spin-spin coupling terms $J_{11} = J_{22} = 0$ and $J_{12} = J_{21} = -1$. The normalized minimizer is ${\bf x}_1 = (1, -1) / \sqrt{2}$ and ${\bf x}_2 = (-1, 1) / \sqrt{2}$, and therefore in this illustration we choose state vectors ${\bf x}_1 = (x, y)$ while setting ${\bf x}_2 = (-1, 1) / \sqrt{2}$. Snapshots of the self-interacting term $H_1$ through the annealing schedule are shown in (a)-(c) for $\alpha = 4$ and $\gamma_i = \gamma = -0.5$ (a), $0.5$ (b), and $1$ (c), illustrating the symmetry-breaking as the gain increases. $H_1 + H_2$ is shown in (d)-(f) which seeks to minimise interaction energy between soft spins; in (f) the minimum is achieved as the the angle between vectors ${\bf x}_1$ and ${\bf x}_2$ is $\pi$; in a more general coupling matrix the vector spins will tend to adjust the angles as to minimise the pairwise interactions. $H_1 + H_2 + H_3$ is shown in (g)-(i) for the same gain parameter as in (a)-(c) and $P(t) =$ $1.0$ (g), $1.5$ (h), and $2.0$ (i). The vector spins are forced to be collinear with the direction that can be identified with the hard Ising spins in (i).}  
    \label{Landscape}
\end{figure}

We illustrate a toy model of VISA for two dimensions in Fig.~(\ref{Landscape}), demonstrating how the energy landscape evolves as the gain $\gamma_i (t)$ and penalty $P(t)$ are evolved, and how the minimizers ${\bf x}^{*}$ and minima of $H_{\rm VISA}$ change during this process.

\section{Numerical Evolution of Vector Ising Spin Annealers}

We consider two popular models to introduce the concept of a scalar-based GBC against which VISA will be benchmarked. (A) Hopfield-Tank (HT) neural networks \cite{hopfield1982neural, hopfield1985neural} have energy landscape given by the Lyapunov function
\begin{equation} \label{HT_Energy}
    E_{\rm HT} = -p(t) \sum_{i = 1}^{N} \int_{0}^{v_i} g^{-1} (x) \md x - \frac{1}{2} \sum_{i, j = 1}^{N} J_{ij} v_i v_j,
\end{equation}
with nonlinear activation function $v_i = g(x_i)$ and real soft-spin variables $x_i (t)$ describing the network state. At any time $t$, the Ising state can be obtained from $x_i$ by associating spins $s_i$ with the sign of the soft-spin variables $s_i = \sgn (x_i)$. The governing equation for HT neural networks is $\dot{x}_i = p(t) x_i + \sum_{j} J_{ij} v_j$ with annealing parameter $p(t)$. This first-order equation can be momentum-enhanced and replaced with the second-order equation leading to Microsoft analogue iterative machine \cite{kalinin2023analog} or Toshiba bifurcation machine \cite{goto2016bifurcation}. We will use these enhancements below. (B) CIM using the degenerate optical parametric oscillators has an energy function
\begin{equation} \label{CIM_Energy}
    E_{\rm CIM} = \frac{1}{4} \sum_{i = 1}^{N} ( p(t) - x_{i}^{2} )^{2} - \frac{\alpha}{2} \sum_{i, j = 1}^{N} J_{ij} x_i x_j,
\end{equation}
where $x_i$, $p(t)$, and $J_{ij}$ represent degenerate optical parametric oscillator quadrature, effective laser pumping power, and coupling strength, respectively. The system evolves as
\begin{equation} \label{CIM_Update_Eq}
    \Dot{x}_i = p(t) x_i - x_{i}^{3} + \alpha \sum_{j} J_{ij} x_j,
\end{equation}
where $\alpha$ is a hyperparameter chosen to maximize solution quality. In the scalar-based GBC models, as the gain $p(t)$ increases from negative values (representing effective losses) to large positive values (large effective gain),  the amplitudes undergo Hopf bifurcation and reach  $x_i = \pm \sqrt{p(T)}$ as $t \rightarrow T,$ $ p(t) = p(T)= {\rm const}$ for $t > T$. At the fixed point, the second term in Eq.~(\ref{HT_Energy}) dominates, which is the target Ising Hamiltonian scaled by $p(T)$. 

The energy landscapes during amplitude bifurcation have many local minima from the increased degrees of freedom of the soft-mode systems. Moreover, the ground state of the soft-mode system may correspond to the excited state of the hard-spin Ising model. At the bifurcation, the system trajectories may get trapped at these minima and could not transition to the correct global minimum at higher gains, especially when global cluster spin flips are required \cite{cummins2023classical}. By enhancing the dimensionality of the energy landscape, VISA may allow the system to evolve from the global minimum of the soft-spin Ising model to the global minimum of the classical hard-spin Ising Hamiltonians, as we now illustrate.

\subsection{$J$-M\"obius Ladder Graph}

First, we examine the Ising Hamiltonian minimization on simple cyclic graphs. These graphs offer analytically solvable benchmarks with distinct and identifiable obstacles in finding ground states. The process of finding Ising ground states on such graphs is significantly influenced by the eigenvalues of the coupling matrix $\mathbf{J}$, especially the relationship between the eigenvectors' component signs and the Ising Hamiltonian's global minimum \cite{cummins2023classical}. For instance, HT networks adjust spin amplitudes to favour the principle eigenvector component signs \cite{kalinin2020complexity}.

\begin{figure}[ht]
\centering
     \includegraphics[width=\columnwidth]{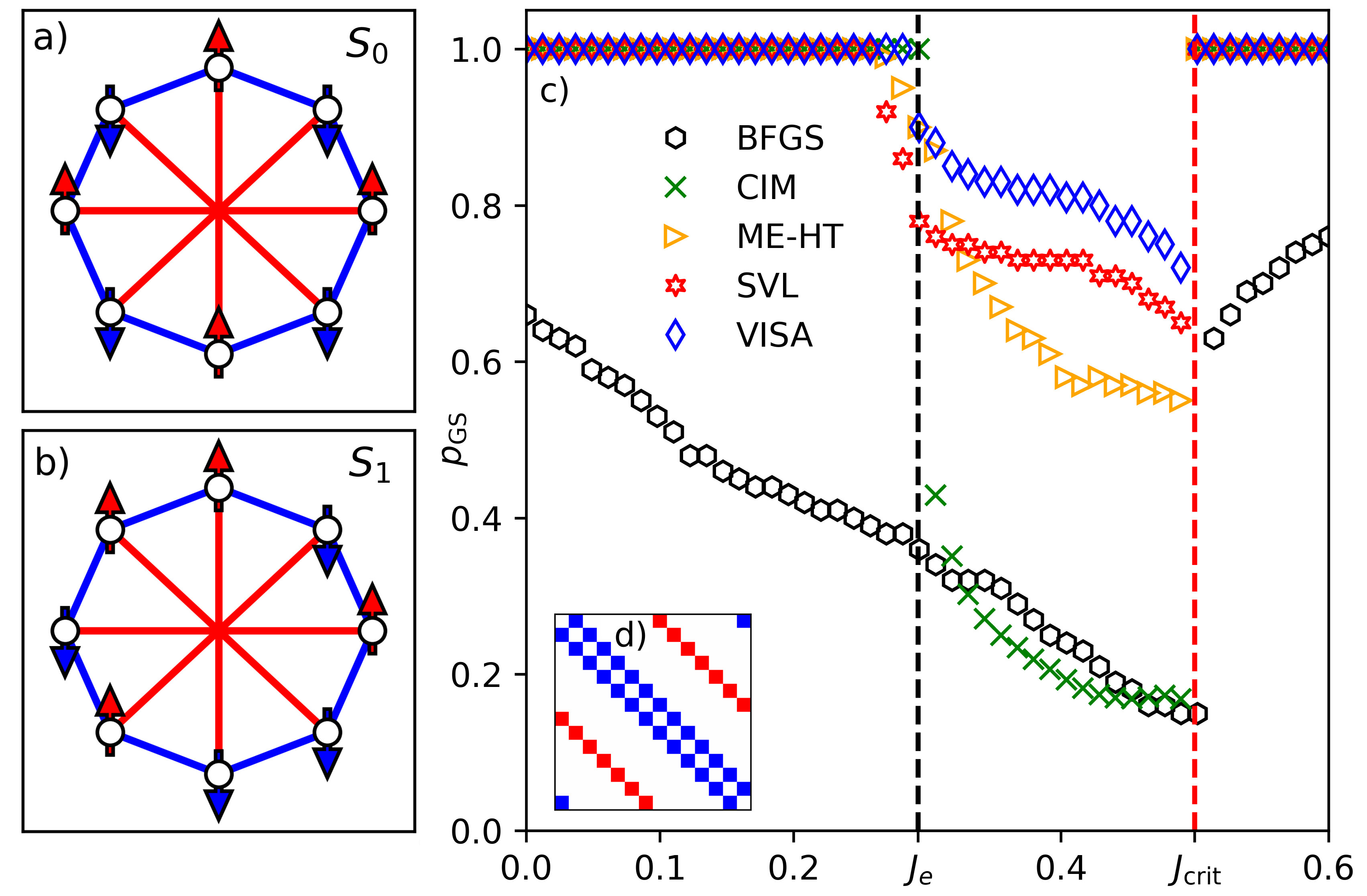}
     \caption{$N = 8$ $J$-M\"obius ladder graphs with circle (blue) and cross-circle (red) couplings for the (a) $S_0$ and (b) $S_1$ configurations. (c) Ground state probability for VISA, SVL, ME-HT, CIM, and BFGS for an $N = 8$ $J$-M\"obius ladder graph with varying cross-circle couplings $J$. One thousand runs are used to calculate the probability of finding the ground state $p_{\rm GS}$ for each value of $J$. Dashed lines corresponding to $J_{\rm e}$ (black) and $J_{\rm crit}$ (red) show the values of $J$ for which the leading eigenvalue changes and the ground state configuration changes, respectively. For VISA, $\varepsilon = 0.02$, $\gamma_i (0) = -0.5$, and $P(t) = \alpha t$. For SVL, $m = 1.0$, $\xi \sim N (\mu = 0, \sigma = 0.1)$, and $\gamma = 0.99$. For momentum-enhanced HT, $m = 1.0$, and $\gamma = 0.99$. For CIM we used $p(t) = (1 - p_0) \tanh (\varepsilon t) + p_0$ with $p_0 = J - 2$, and $\varepsilon = 0.003$. For each value of $J$, optimal values of $\alpha$ and $\beta(0)$ are chosen based on sets of preliminary runs in which they are varied. Inset (d) illustrates the structure of M\"obius ladder coupling matrices ${\bf J}$ with $J_{ij} = 0$ (white), $-1$ (blue), $-J$ (red).}
    \label{Probability_Results}
\end{figure}

We start by using a M\"obius ladder-type graph, a cyclic structure with an even number of vertices $N$ arranged in a ring, featuring variable couplings between adjacent and opposite vertices. To explore non-trivial ground states, we introduce equal antiferromagnetic couplings ($J_{ij} = -1$) among nearest neighbors and variable cross-ring antiferromagnetic couplings ($J_{ij} = -J$). We refer to the M\"obius graphs with these couplings as {\it $J$-M\"obius graph}. Following Ref.~\cite{cummins2023classical}, we define $S_0$ as the alternating spin state around the ring and $S_1$ as the state with spins alternating except at two opposite ring points with frustrated spins, as depicted in Fig.~(\ref{Probability_Results})(a) and (b). When $N/2$ is even, the energies and principal eigenvalues of $S_0$ and $S_1$ intersect at $J_{\rm crit} \equiv 4/N$ and $J_{\rm e} \equiv 1 - \cos (2 \pi / N)$, respectively. In the range $J_{\rm e} < J < J_{\rm crit}$, $S_0$'s largest eigenvalues are smaller than those of $S_1$, with an eigenvalue gap $\Delta = 2 \cos (2 \pi / N) + 2J - 2$, even though $S_0$ represents the ground state with lower energy. The dynamics of semi-classical soft-spin models, including these considerations, are juxtaposed with quantum annealing approaches in the minimization of the Ising Hamiltonian on $J$-M\"obius ladder graphs, as detailed in Ref.~\cite{cummins2023classical}.

The presence of amplitude heterogeneities enables the soft-spin model to acquire and maintain its state achieved at the bifurcation even when it diverges from the classical hard-spin Ising Hamiltonian ground state, potentially complicating the optimization process when $J_{\rm e} < J < J_{\rm crit}$. To mitigate this issue, the manifold reduction CIM (MR-CIM) technique was developed, incorporating an additional feedback mechanism to regulate soft spin amplitudes, ensuring they remain close to their mean value \cite{cummins2023classical}. The amplitude adjustment after each update at time step $t$ is governed by
\begin{equation} \label{MR-CIM}
    x_i \rightarrow (1 - \delta) x_i + \delta \frac{R({\bf x}) x_i}{|x_i|},
\end{equation}
where $0 < \delta < 1$. This adjustment draws the spins nearer to the mean, with the average defined by the squared radius of the quadrature $R(\mathbf{x}) = \sum_i x_i^2 / N$, thus aligning the local and global minima of the soft and hard spin models more closely. Comparative analyses between various CIM modes and quantum annealing have been conducted, demonstrating that, despite quantum annealing's ability to utilize quantum entanglement and inter-spin correlations to identify ground states, it exhibits heightened sensitivity to diminishing energy gaps near $J_{\rm crit}$. Consequently, quantum annealing demands longer annealing schedules to accurately determine ground state solutions as $J$ nears $J_{\rm crit}$ \cite{cummins2023classical}.

\begin{figure}[ht]
\centering
     \includegraphics[width=\columnwidth]{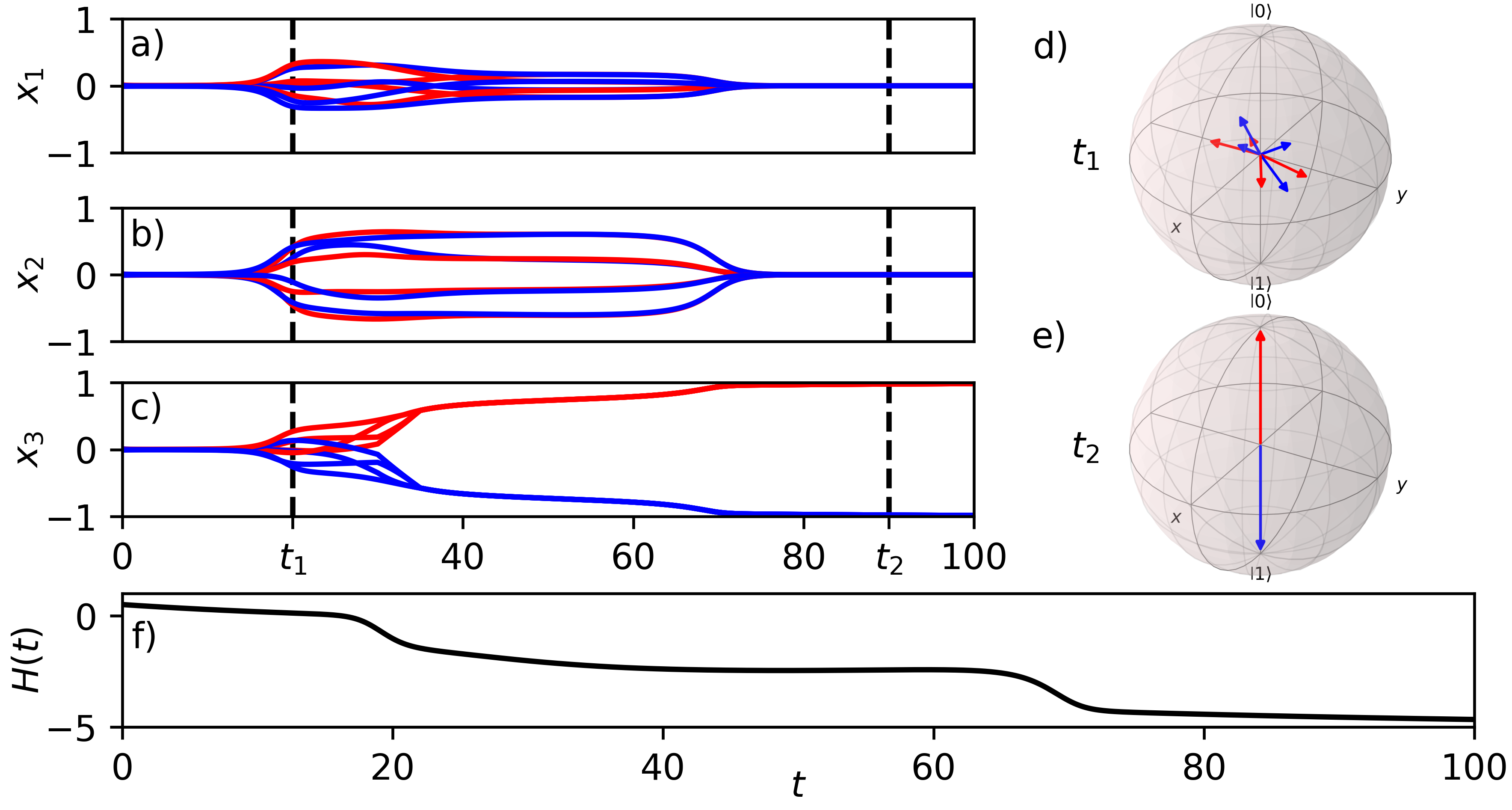}
     \caption{Evolution of the VISA model in an $N = 8$ $J$-M\"obius ladder network with coupling strength $J = 0.4$, as governed by Eq.~(\ref{DA_Evo_Eq}). This figure illustrates the dynamics of VISA's soft-spin components: $x_1 = x_{1i}$ (a), $x_2 = x_{2i}$ (b), and $x_3 = x_{3i}$ (c), with vertical dashed lines indicating the snapshots at times $t_1$ and $t_2$. Panels (d) and (e) depict the corresponding spin vectors, highlighting the orientation at $t_1$--influenced by $H_1 + H_2$--and at $t_2$, where spins align in a unified direction due to a high $P(t_2)$, revealing the hard-spin Ising Hamiltonian's global minimum. The final orientation of the vectors at $t_2$ is spontaneous; thus, the vectors are adjusted such that the $z$-axis aligns with this direction. Panel (f) shows the trajectory of the VISA Hamiltonian, $H_{\rm VISA}$, over time as it converges to the ground state. The numerical integration of Eq.~(\ref{DA_Evo_Eq}) utilized the following parameters $\varepsilon = 0.03$, $\gamma_i (0) = -0.5$, $P(t) = t/200$, $\alpha = 4$, and initial conditions $x_{ij} (0)$ uniformly distributed within $[-0.01, 0.01]$. The fourth-order Runge-Kutta method with fixed time step $\Delta t = 0.1$ was employed for solving Eq.~(\ref{DA_Evo_Eq}).} 
    \label{DA}
\end{figure}

Figure~(\ref{DA}) illustrates the application of VISA to a $J$-M\"obius ladder graph instance characterized by cross-circle couplings where $J_{\rm e} < J < J_{\rm crit}$. The continuous spin components experience an Aharonov-Hopf bifurcation during the process, exploiting the minimal energy barriers inherent in soft-spin models. We particularly focus on the intermediary time $t_1$ following the bifurcation, noting that at this time, the spin amplitudes have yet to achieve unit magnitude, and the spins are not collinear. By a later time $t_2 > t_1$, the spin vectors stabilize as the minimization of the VISA Hamiltonian progresses. These soft-spin vectors then exhibit Ising spin characteristics, including unit magnitude and (anti-)parallel alignment, while the coupling term $H_2$ guarantees identifying the target Ising Hamiltonian's ground state.

\begin{figure}[ht]
\centering
     \includegraphics[width=\columnwidth]{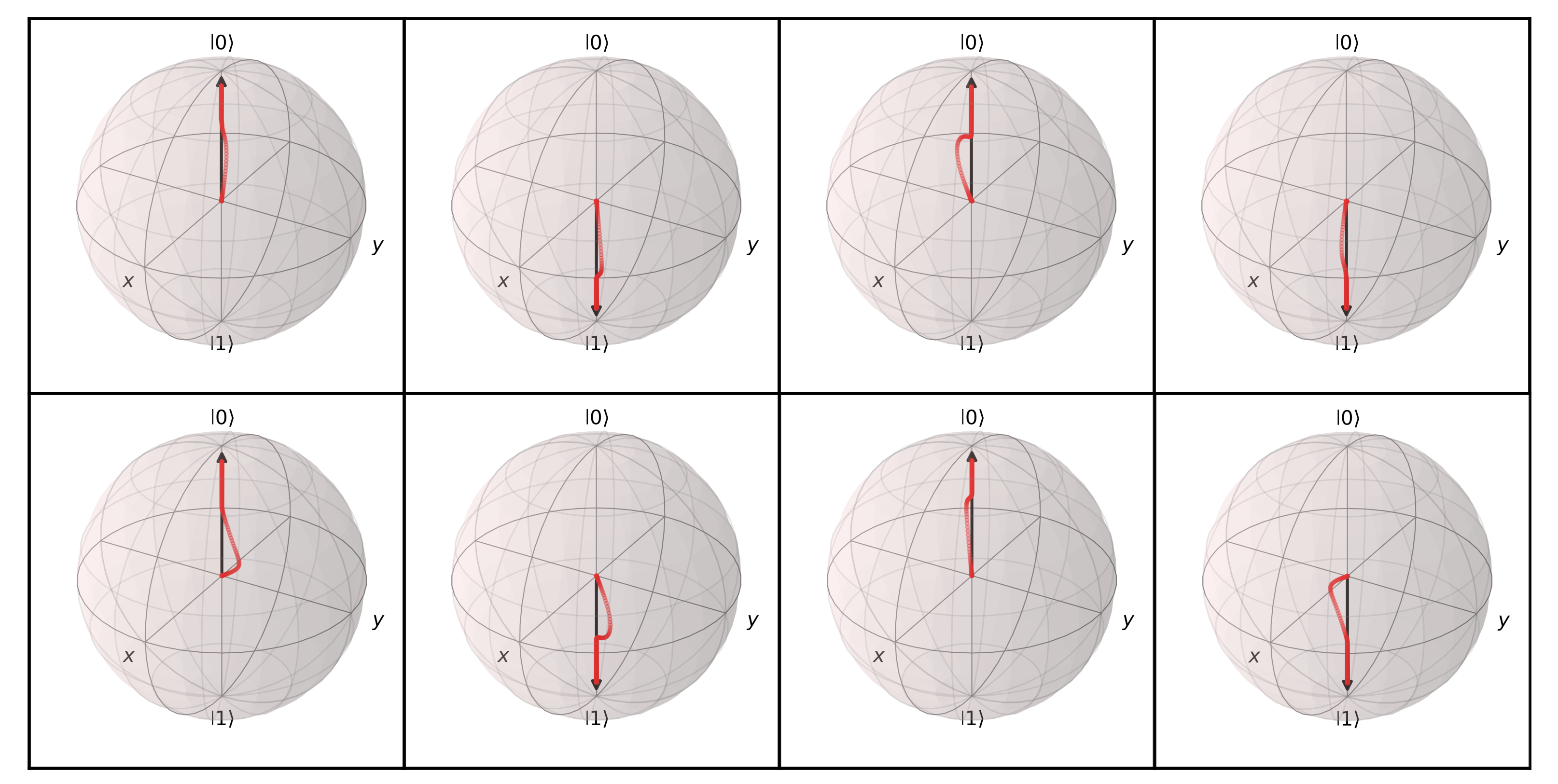}
     \caption{Evolution of VISA spins for an $N = 8$ $J$-M\"obius graph with coupling constant $J = 0.4$. The spins reach final states adhering to the Ising model's collinear and unit magnitude constraints. The spontaneous orientation of vectors at time $t_2$ leads to an adjustment of the spin vectors, aligning the $z$-axis with the resultant direction. Parameters used are $\varepsilon = 0.04$, $\gamma_i(0) = -0.5$, $P(t) = t/25$, $\alpha = 4$, and initial conditions $x_{pi}(0)$ uniformly distributed within $[-0.01, 0.01]$.}
    \label{Spheres}
\end{figure}

In Fig.~(\ref{Spheres}), we depict typical trajectories of VISA vectors on Bloch spheres, drawing a parallel to similar quantum representations. We orient the spontaneously aligning spins along the $z$-axis to facilitate visualisation. With a fixed interaction strength where $J_{\rm e} < J < J_{\rm crit}$ and considering $N = 8$, we demonstrate the bifurcation dynamics originating from the center. This visualization shows how the three-dimensional nature of VISA allows spins to traverse paths that connect various minima, ultimately converging on the ground state $S_0$. The final vector states are indicated by black arrows, while red points mark the terminal positions of spin vectors at sequential time steps $t$. These trajectories elucidate the process by which spins increasingly satisfy collinearity and unit magnitude constraints through external annealing of $P(t)$ and the feedback mechanism on $\gamma_i$, steering the spin magnitudes to the threshold value of $|{\bf x}_i| = 1$.

\begin{figure}[ht]
\centering
     \includegraphics[width=\columnwidth]{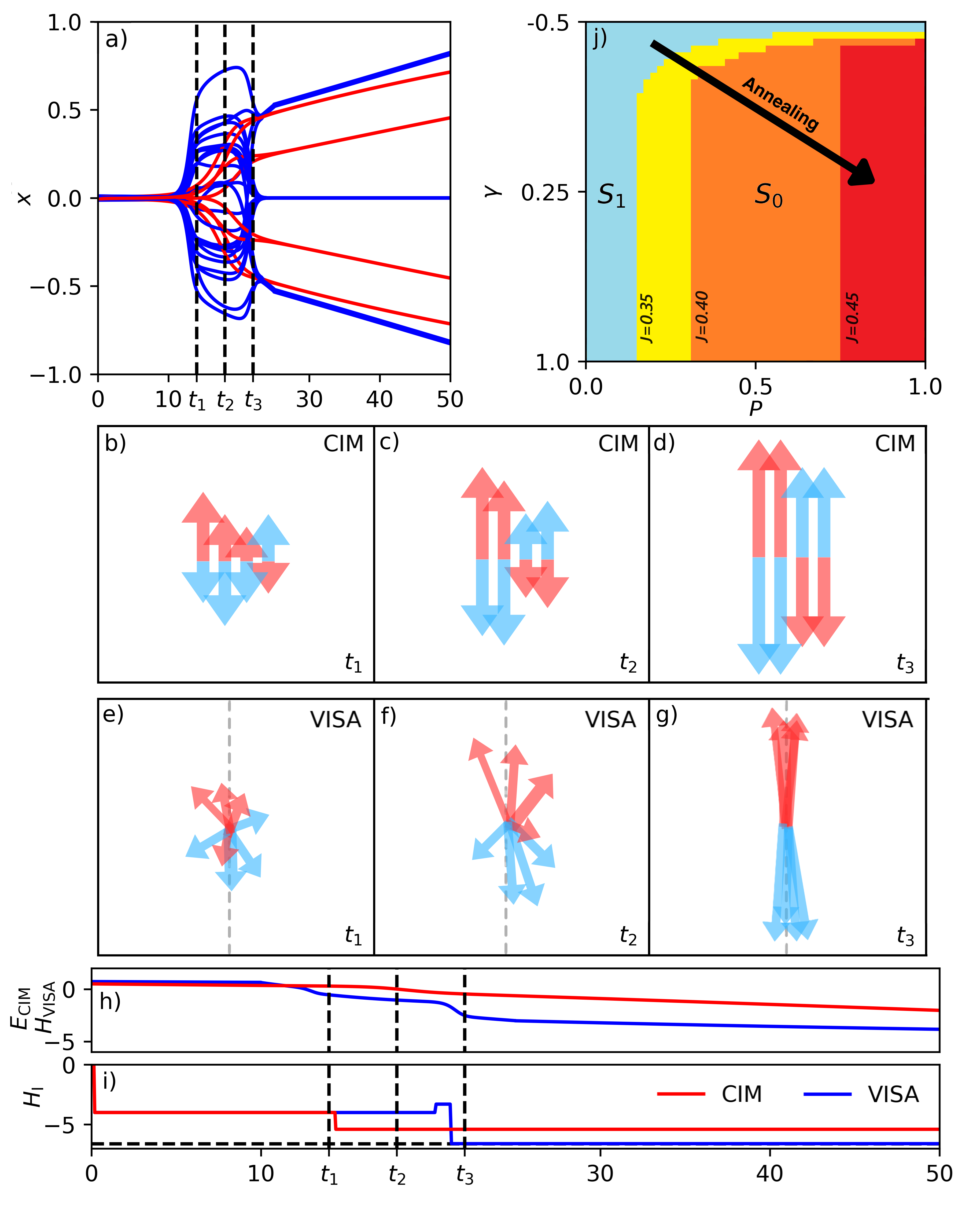}
     \caption{Panel (a) compares the trajectory evolution of VISA (blue) and CIM (red) on an $N = 8$ $J$-M\"obius ladder graph with a coupling constant $J = 0.35$. Here, VISA successfully reaches the ground state $S_0$, while CIM remains in an excited state $S_1$. Both systems start from the same initial conditions, with CIM beginning at $x_i (0) = a$ and VISA at ${\bf x}_i (0) = (a, b, b)$, where $a$ and $b$ are uniformly chosen from the ranges $[-0.01, 0.01]$ and $[-0.0001, 0.0001]$ respectively. Panels (b)-(d) for CIM and (e)-(g) for VISA show the spin states at three different times $t_1$, $t_2$, and $t_3$, marked by vertical dashed lines in panel (a), illustrating CIM's inability to overcome the energy barrier to reach the ground state, in contrast to VISA's effective navigation in three-dimensional space. Panel (h) compares the VISA Hamiltonian $H_{\rm VISA}$ against the CIM energy $E_{\rm CIM}$ as calculated from Eq.~(\ref{CIM_Energy}), while panel (i) focuses on the Ising energy, with the ground state $S_0$ indicated by a black dashed line. Lastly, panel (j) outlines the regions in the $\gamma - P$ space that correspond to different global minima of $H_{\rm VISA}$ for $N = 8$, showing $S_1$ in blue and $S_0$ in varying colors (yellow, orange, red) for $J$ values of $0.35$, $0.40$, and $0.45$, respectively.}
    \label{Comparison}
\end{figure}

In Fig.~(\ref{Comparison}), we compare VISA with CIM under equivalent starting conditions, highlighting that while CIM cannot find the ground state within the range $J_{\rm e} < J < J_{\rm crit}$, VISA excels by leveraging the dynamics of spins in three dimensions to identify the global minimum. Soft-spins contribute to this success in both models by facilitating the escape from local minima through reduced energy barriers, a feature not present in hard-spin models. Unlike CIM, constrained by the principal eigenvalue of ${\bf J}$ to an excited state, VISA uses its multidimensional advantage to bridge minima unreachable by CIM's one-dimensional approach. Additionally, VISA's mechanism allows for spin flipping at energy costs lower than that required by CIM, thanks to its ability to navigate through three-dimensional space to find the most energy-efficient path to the global minimum. Figures~(\ref{Comparison})(a)-(i) depict how VISA effectively minimizes Ising energy, contrasting with CIM's stagnation in an excited state. For $J_{\rm e} < J < J_{\rm crit}$, and as the gain $\gamma_i$ and penalty $P(t)$ increase, VISA goes towards the ground state $S_0$, aligning with the lowest energy state indicated by the hard-spin Ising Hamiltonian, as shown in Fig.~(\ref{Comparison})(j). Nonetheless, it's important to note that increasing spin amplitudes too fast may also inadvertently heighten energy barriers, potentially impeding state transitions. Further insight into the structure of the VISA energy landscape and its comparison with the energy landscape of the scalar models can be gained by analysing the critical points (see Supplementary Information). The VISA energy landscape shows much-diminished energy barriers.

We compare VISA to two second-order scalar networks methods, namely (i) momentum-enhanced Hopfield-Tank (ME-HT), and (ii) spin-vector Langevin. With rigorous hyper-parameter exploration phases, ME-HT outperforms parallel tempering, simulated annealing, and commercial solver Gurobi at various QUBO benchmarks \cite{kalinin2023analog}. The ME-HT governing equation is
\begin{equation} \label{ME-HT}
    m \Ddot{x}_i + \gamma \Dot{x}_i - \beta (t) x_i - \alpha \frac{\partial H ( g (\mathbf{x}) )}{\partial x_i} = 0,
\end{equation}
with effective mass $m$, momentum term $\gamma \in [0, 1)$, and hyperparameters $\alpha$ and $\beta (t)$, the latter of which undergoes an annealing protocol given by $\beta (t) = \beta_0 (1 - t/T)$. Equation (\ref{ME-HT}) combines gradient descent with annealed non-conservative dissipation, and the addition of momentum distinguishes it from regular first-order HT networks with energy landscape Eq.~(\ref{HT_Energy}). Momentum, or the heavy-ball method, aims to overcome the pitfalls of pathological curvature in deep learning and accelerates standard gradient descender optimizers \cite{orvieto2019shadowing, saab2022adaptive}. 

VISA can be further compared and contrasted with the spin-vector Langevin (SVL) model that was proposed as a classical analogue of quantum annealing description using stochastic Langevin time evolution governed by the fluctuation-dissipation theorem \cite{subires2022benchmarking}. SVL is based on the time-dependent Hamiltonian used in quantum annealing $H(t) = A(t) H_0 + B(t) H_P$, where initial Hamiltonian $H_0 = \sum_i \sigma_{i}^{x}$, and problem Hamiltonian $H_P = - \sum_{i, j} J_{ij} \sigma_{i}^{z} \sigma_{j}^{z}$, with Pauli operator ${\bf \sigma}_i$ acting on the $i$-th variable. Real annealing functions satisfy boundary conditions $A(0) = B(T) = 1$ and $A(T) = B(0) = 0$, where $T$ is the temporal length of the annealing schedule. If the rate of change of the functions is slow enough, the system stays in the ground state of the instantaneous Hamiltonian so that the Ising Hamiltonian is minimised at $t = T$. Quantum annealing has shown competitive results in quadratic unconstrained binary optimization (QUBO) problems such as subset sum, vertex cover, graph coloring, and travelling salesperson \cite{jiang2022solving}. The SVL model replaces Pauli operators with real functions of continuous angle $\sigma_{i}^{z} \rightarrow \sin \theta_i$, $\sigma_{i}^{x} \rightarrow \cos \theta_i$, and is therefore a classical annealing Hamiltonian using continuous-valued spins $\sin \theta_i$. SVL dynamics is described by a system of coupled stochastic equations
\begin{equation} \label{SVL_Eq}
    m \Ddot{\theta}_i + \gamma \Dot{\theta}_i + \alpha \frac{\partial H (\mathbf{\theta})}{\partial \theta_i} + \xi_i (t) = 0,
\end{equation}
where $m$ is the effective mass, $\gamma$ is the damping constant, $\alpha$ is a hyperparameter, and $\xi_i (t)$ is an iid Gaussian noise. For long annealing times, the minima of HSIHs are obtained through the transformation $s_i = \sgn ( \sin \theta_i )$.
The gradient term in Eq.~(\ref{SVL_Eq}) is
\begin{equation}
    \frac{\partial H (\mathbf{\theta})}{\partial \theta_i} = - B(t) \sum_{j = 1}^{N} J_{ij} \cos \theta_i \sin \theta_j + A(t) \sin \theta_i,
\end{equation}
which in conjunction with fluctuation-dissipation relations $\langle \xi_i (t) \rangle = 0$ and $\langle \xi_i (t) \xi_j (t ') \rangle = \delta_{ij} \delta (t - t ')$, give $2N$ stochastic differential equations: $\md \theta_i = (p_i / m) \md t$ and
\begin{equation}
    \md p_i = \Big( \frac{\partial H (\mathbf{\theta})}{\partial \theta_i} + \frac{\gamma}{m} p_i \Big) \md t + \md W_i,
\end{equation}
where $\md W_i$ represents a real-valued continuous-time stochastic Wiener process \cite{subires2022benchmarking}. The characteristic amplitude bifurcation of scalar spins according to ME-HT and SVL are given in Supplementary Information.

The key feature distinguishing VISA from the discussed models lies in its novel gain-based and dimensionality annealing strategy applied  across multiple dimensions.  In the analysis of $J$-M\"obius ladder graphs, as shown in Fig.~(\ref{Probability_Results}), we compute the ground state probability for VISA alongside SVL, ME-HT, CIM, and the Broyden-Fletcher-Goldfarb-Shanno (BFGS) algorithm \cite{bfgsnote}. Within the range $J_{\rm e} < J < J_{\rm crit}$, VISA consistently identifies the ground state $S_0$ with a higher probability $p_{\rm GS}$ compared to the other models. 

\subsection{$J-G$ Cyclic Graphs}

\begin{figure}[ht]
\centering
     \includegraphics[width=\columnwidth]{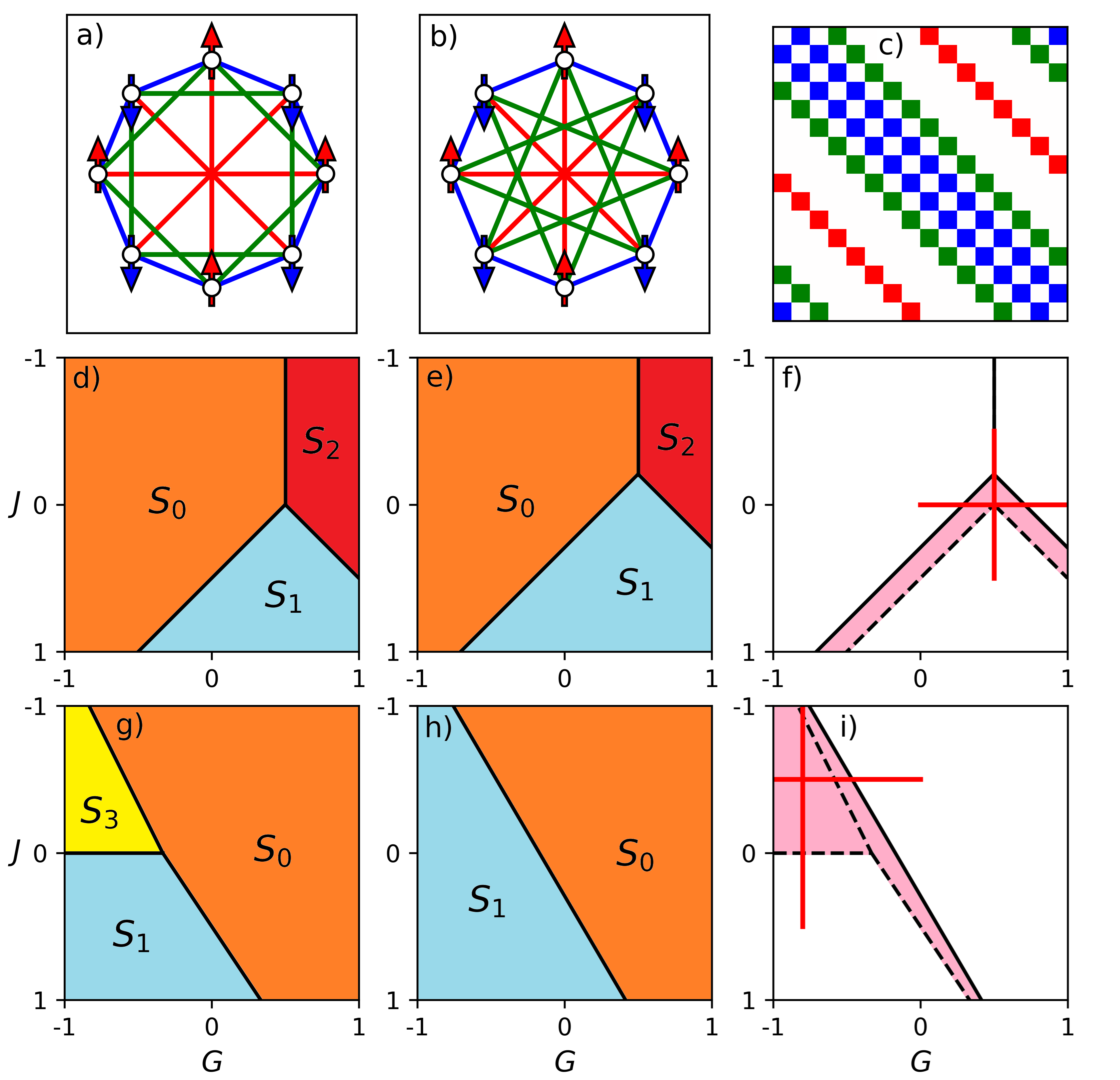}
     \caption{$J-G$ cyclic graphs for $N = 8$ with couplings (a) $k = 2$, and (b) $k = 3$. Panel (c) depicts the matrix structure ${\bf J}$ for $k = 3$, where the color coding represents $J_{ij} = 0$ (white), $-1$ (blue), $-J$ (red), and $-G$ (green). The ground states and leading eigenvalue states for $k = 2$ are displayed in panels (d) and (e) within the $J - G$ space, respectively. Panel (f) delineates the energy and eigenvalue boundaries, with the pink region indicating discrepancies between ground and maximal eigenvalue states. The sequence from panels (g) to (i) extends this analysis to $k = 3$, highlighting the ground state probabilities across easy-hard transitions in the $J - G$ space, indicated by solid red lines in panels (f) and (i).}
    \label{New_Mobius}
\end{figure}

Next we consider the {\it $J-G$ cyclic graphs} that are a variant of $J$-M\"obius ladder graphs that include additional couplings, connecting each vertex $i$ to vertices $i \pm k$ with a weight of $-G$, where $1 < k < N/2$. This modification aims to explore more complex interaction patterns and their impact on the system's ground state behavior and eigenvalue distributions, particularly focusing on how these factors evolve in optimization and the search for ground states in varied graph structures. The interaction strength range is expanded to accommodate both ferromagnetic and antiferromagnetic interactions, with $-1 \leq J, G \leq 1$. This yields a 5-regular circulant graph characterized by weights $J_{ij} \in \{-1, -J, -G \}$, depicted in Fig.~(\ref{New_Mobius}). Cyclic graphs maintain their local and global topological properties under rotational transformations, encapsulating all connectivity information within any row of ${\bf J}$. By selecting the first row $J_{1,j}$, we compute eigenvalues as $\lambda_n = \sum_{j = 1}^{N} J_{1, j} \cos \left[ 2 \pi n (j - 1) / N \right]$, leading to $\lambda_n = - 2 \cos (2 \pi n / N) - J (-1)^{n} - 2 G \cos (2 \pi k n / N)$ \cite{gancio2022critical}. We then deduce boundaries by observing eigenvalues in the $J - G$ plane, identifying where leading eigenvalues and corresponding eigenvectors change their correspondence with the ground and excited states. The ground state boundaries are identified within the $J - G$ space in the Supplementary Information. Figure (\ref{New_Mobius}) depicts regions where the global minimum diverges from the hypercube corner of the projected eigenvector with the highest eigenvalue.

\begin{figure}[ht]
\centering
     \includegraphics[width=\columnwidth]{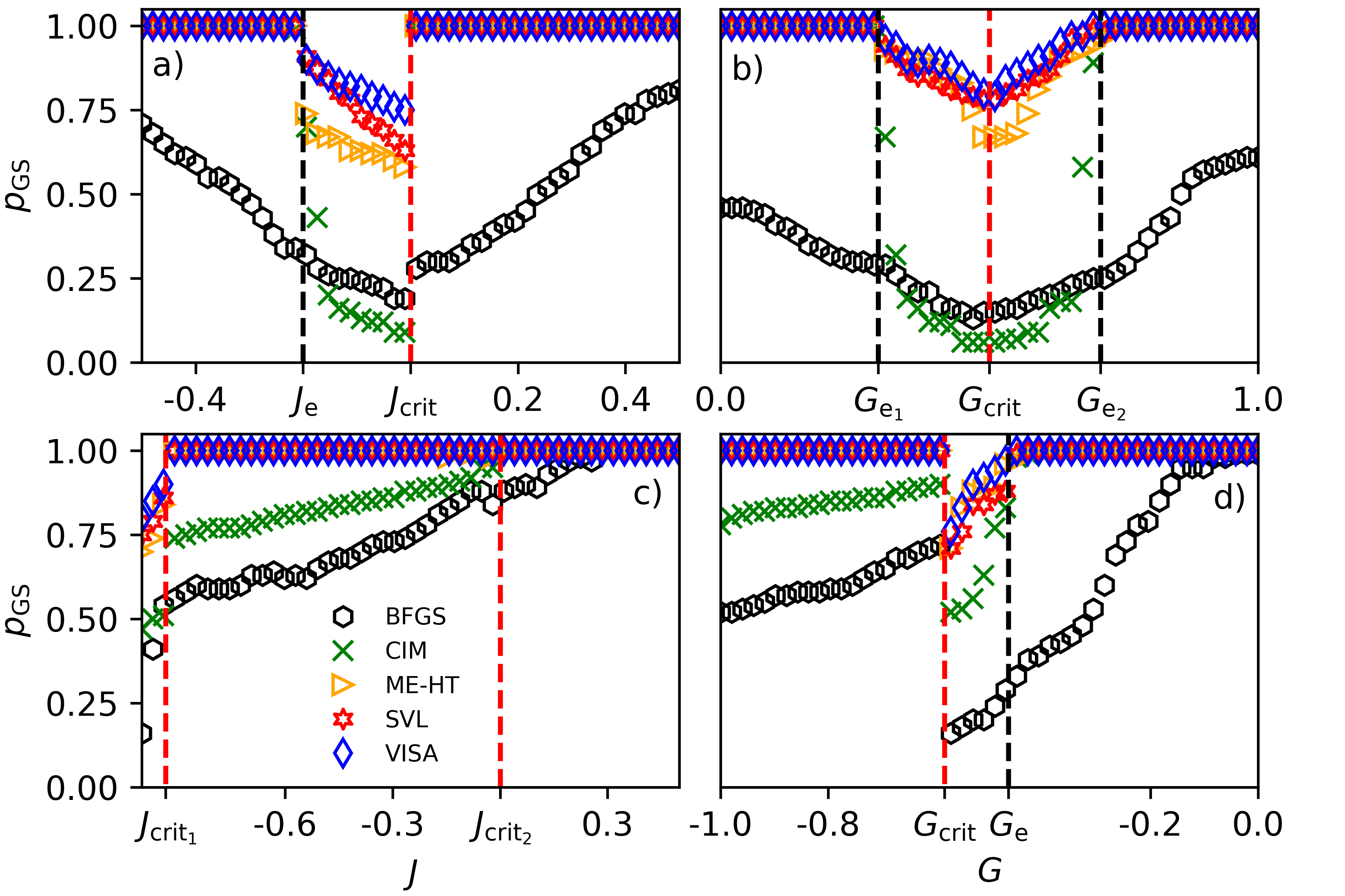}
     \caption{Ground state probability for VISA, SVL, ME-HT, CIM, and BFGS for $N = 8$ $J-G$ cyclic graphs with varying cross-circle couplings $J$ and $G$. (a) $k = 2$, $J \in [-0.5, 0.5]$, and $G = 0.5$. (b) $k = 2$, $J = 0.5$, and $G \in [0, 1]$. (c) $k = 3$, $J \in [-1, 0.5]$, and $G = -0.8$. (d) $k = 3$, $J = -0.5$, and $G \in [-1, 0]$. One thousand runs are used to calculate the probability of finding the ground state $p_{\rm GS}$ for each value of $J$. Dashed red and black lines show where the ground state energies and leading eigenvalues change, respectively. For VISA, $\varepsilon = 0.04$, $\gamma_i (0) = -0.5$, and $P(t) = \alpha t$. For SVL, $m = 1.0$, $\xi \sim N (\mu = 0, \sigma = 0.1)$, and $\gamma = 0.99$. For momentum-enhanced HT, $m = 1.0$, and $\gamma = 0.99$. The general annealing protocol for CIM is used with $p_0 = J - 2$ and $\varepsilon = 0.003$. For each value of $J$, optimal values of $\alpha$ and $\beta(0)$ are chosen based on sets of preliminary runs in which they are varied.}
    \label{New_Mobius_Probability}
\end{figure}

We investigate ground state probabilities $p_{\rm GS}$ over transition regions in these  graphs with both $J$ and $G$ cross-circle couplings. We choose values in $J - G$ parameter space that demonstrate transitions between regions in which the eigenvector corresponding to the leading eigenvalue does not match the ground state solution (exact calculations of these regions are presented in Supp.~Inf.). Ground state probabilities for VISA, SVL, ME-HT, CIM, and BFGS are illustrated in Fig.~(\ref{New_Mobius_Probability}). Four cases are analyzed, split into two sets: $k = 2$ and $k = 3$. We consider perpendicular lines in the two-dimensional $J - G$ space for each set. Specifically, we vary (fix) $J$ and fix (vary) $G$. For $k = 2$ and $G = 0.5$, an easy-hard-easy transition emerges as $J$ increases, akin to $J$-M\"obius ladder previously studied. Indeed, for $J_{\rm e} < J < J_{\rm crit}$, $p_{GS}$ decreases as the eigenvalue gap $\Delta$ increases. If, instead, we fix $J = -0.5$, a transition occurs, centred at the change between ground states given by $G_{\rm crit} = 0.5$. Here, the hard region is bounded by eigenvalue crossing points $G_{\rm e_1} = 1 - \sqrt{2}/2$ and $G_{\rm e_2} = 1 / \sqrt{2}$. For $k = 3$, $p_{\rm GS}$ is less sensitive to the magnitude of $\Delta$ for regions where $S_{3}$ is the ground state. This is due to the proximity between the leading eigenvalue state $S_{1}$ and ground solution $S_{3}$ in topological spin space. More precisely, the transformation from $S_{1}$ to $S_{3}$ requires only a single spin flip, representing a nominal energy barrier for soft-spin models. Therefore, hardness in $J-G$ cyclic graphs  derives not just from the eigenvalue gap magnitude but additionally from the distance between hypercube corners of the ground and leading eigenvalue states.

\subsection{Random Graphs}

\begin{figure}[ht]
\centering
     \includegraphics[width=\columnwidth]{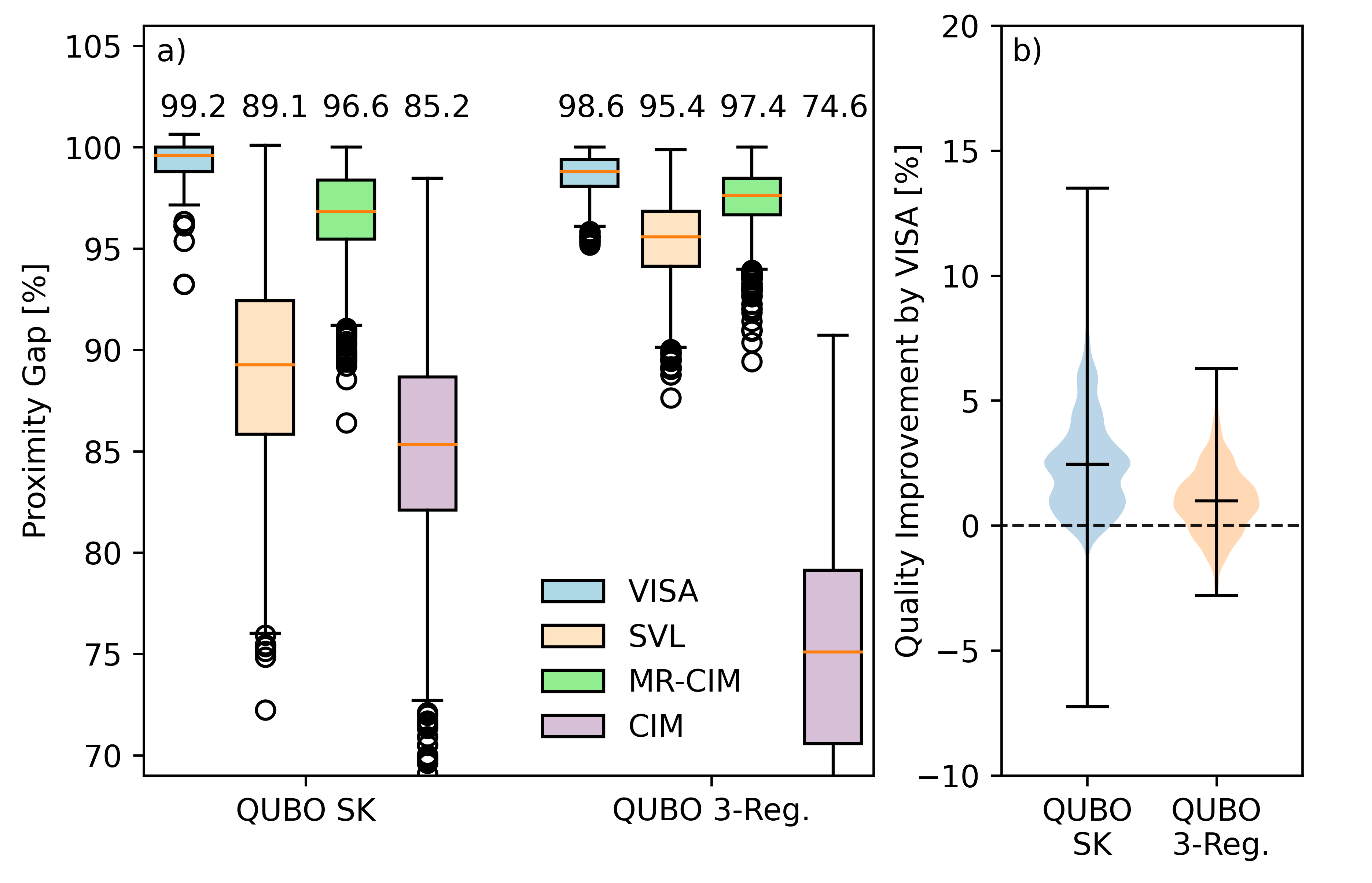}
     \caption{(a) The proximity gap, defined as the ratio of the found objective to the Gurobi objective, for VISA, SVL, manifold reduction CIM, and CIM methods. Gurobi was given a 30-second time limit for each of the 100 problem instances with size $N = 100$. The instances are divided equally between two graph topologies: dense, fully connected, and sparse three-regular graphs. In both cases, the matrix weight elements are drawn from the Gaussian distribution with zero mean and unit variance, resulting in instances belonging to the Sherrington-Kirkpatrick and weighted 3-regular Max-Cut problems. (b) Violin plots demonstrate the distribution of VISA's quality improvement performance compared to the best solution found by competing methods \{SVL, MR-CIM, CIM\} across the SK and 3-Regular QUBO benchmarks.}
    \label{Statistic_Results}
\end{figure}

We extend VISA's evaluation to QUBO instances renowned for their computational intensity as they scale: dense fully connected graphs and sparse three-regular graphs with customarily distributed random couplings $J_{ij}$, representing Sherrington-Kirkpatrick (SK) and weighted three-regular Max-Cut problems, respectively. These models are pivotal for benchmarking physical simulators \cite{haribara2017performance, hamerly2019experimental, harrigan2021quantum, bohm2019poor} and are categorized under the NP-hard complexity class \cite{arora2005non, lucas2014ising}. Unlike cyclic graphs, these instances do not have analytically known ground states, necessitating using the Gurobi optimization suited for ground state estimations. Gurobi employs advanced pre-processing and heuristic enhancements to expedite branch-and-bound algorithms \cite{gurobi}. Figure~(\ref{Statistic_Results})(a) compares the ground state approximations achieved by Gurobi with those by VISA for $N = 100$ SK and weighted three-regular Max-Cut graphs. Alongside, we include comparisons with SVL, MR-CIM, and CIM, where MR-CIM and CIM adapt their laser intensities following a general pumping scheme $p(t) = (1 - p_0) \tanh(\varepsilon t) + p_0$. MR-CIM further applies an additional feedback mechanism as per Eq.~(\ref{MR-CIM}) on top of Eq.~(\ref{CIM_Update_Eq}), controlling soft-spin amplitudes and the dimensionality landscape. We define the quality improvement of VISA over another method $X$ in terms of objective values $O$ as $(O_{\rm VISA} - O_{X}) / O_{\rm VISA}$, showcasing these metrics for SK and three-regular problems in Fig.~(\ref{Statistic_Results})(b), where $X$ represents the best-performing competing method for each instance.

\section{Conclusions}

This paper introduces the Vector Ising Spin Annealer (VISA). This model capitalizes on the advantages of multidimensional spin systems, gain-based operation and soft-spin annealing techniques to optimize Ising Hamiltonians on various graph structures. VISA distinguishes itself by enabling more efficient navigation through the solution space, enhancing spin mobility in higher-dimensional spaces, and providing a robust framework for connecting local minima and reducing energy barriers.

A key focus of VISA is its ability to recover ground states effectively, even in scenarios where these states do not correspond to the principal eigenvector of the coupling matrix. The model's performance was numerically evaluated against other methods, demonstrating its superior ability to find ground states across different graph types and complex QUBO instances. Thus, it highlights its potential to address NP-hard problems.

Future research could explore the role of defects in spin models, such as topological defects, domain walls, and vortex rings and their role in achieving the ground state during gain-based operation. Vortices may exhibit more efficient annihilation properties in higher-dimensional systems, such as those utilized by VISA. This is potentially due to the additional spatial degrees of freedom, which could facilitate the merging or cancellation of vortices and anti-vortices, a phenomenon less constrained than in two-dimensional spaces. This enhanced annihilation could lead to a smoother energy landscape, aiding the system in avoiding local minima and more effectively converging to the ground state.

By leveraging multidimensional spins, VISA opens new avenues for developing analogue optimization machines, potentially incorporating quantum effects to enhance computational capabilities. The prospects of applying dimensionality annealing techniques in optical-based Ising machines suggest an exciting future for speed-of-light computation and accurate ground state recovery, marking a significant advancement in optimization technologies.

\section{Acknowledgements}

J.S.C.~acknowledges the PhD support from the EPSRC. N.G.B.~acknowledges the support from the Julian Schwinger Foundation Grant No.~JSF-19-02-0005, the HORIZON EIC-2022-PATHFINDERCHALLENGES-01 HEISINGBERG project 101114978, and Weizmann-UK Make Connection grant 142568. Both authors are grateful to Dr Fernando G\'omez-Ruiz for valuable discussions.

\section{Supplementary Information}

\subsection{Critical Points of VISA Hamiltonian}

\begin{figure}[ht]
\centering
     \includegraphics[width=\columnwidth]{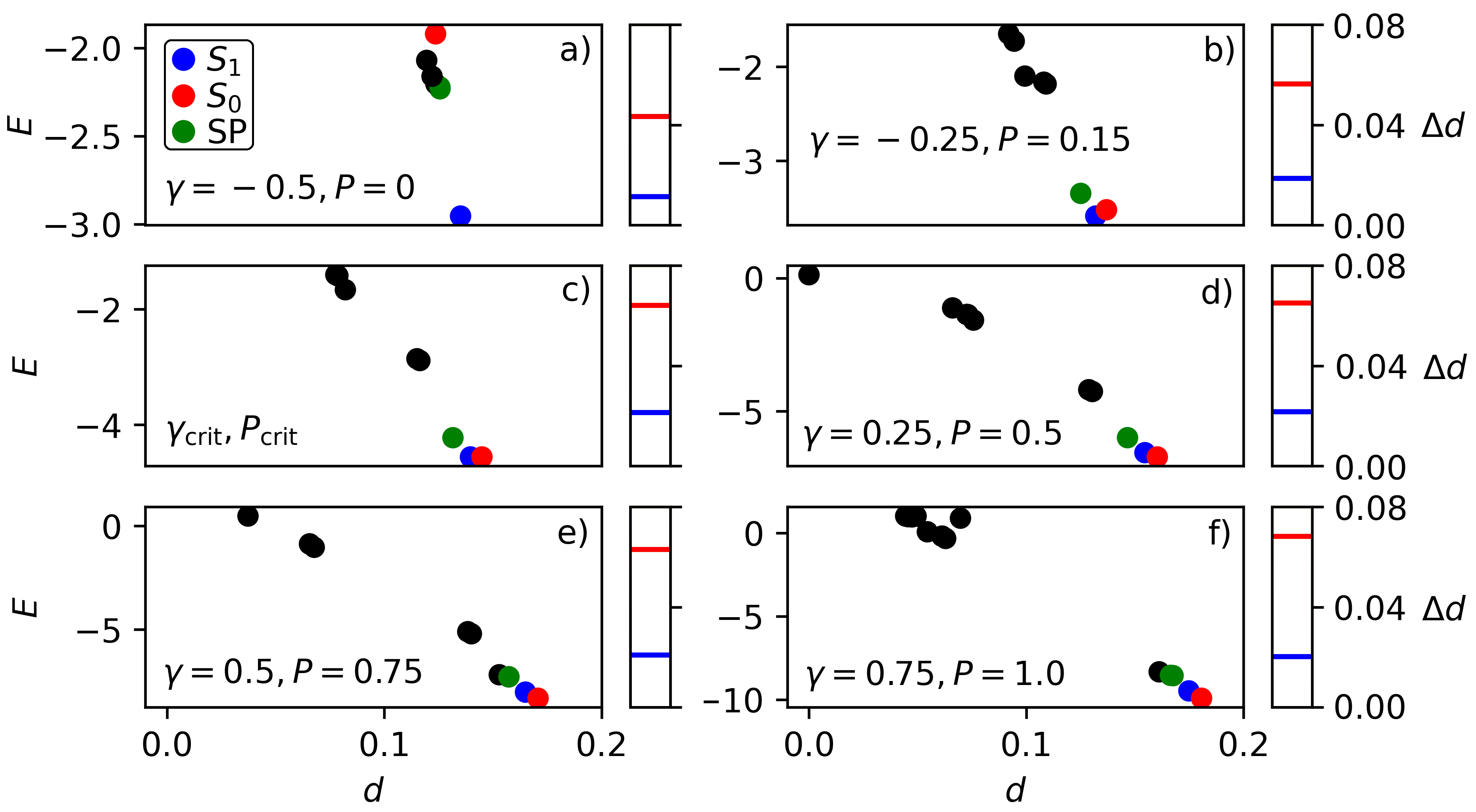}
     \caption{Critical points of $H_{\rm VISA}$ for $N = 8$ $J$-M\"obius ladder graph with $J = 0.4$. Each point is characterized by its energy $E$ and Euclidean distance from the origin $d$. The number of soft spins normalizes distance $d$. Blue and red circles denote minima for states $S_1$ and $S_0$, respectively, while black circles indicate other minima and saddle points of $H_{\rm VISA}$. The closest saddle point ({\rm SP}) to excited state $S_1$ is highlighted in green. Progressing from figures (a) to (f), we systematically increase the gain $\gamma_i = \gamma$ for all $i$ and the penalty $P$, following a typical annealing schedule. Notably, at (c), with $\gamma = -0.087$ and $P = 0.32$, we reach a critical juncture that delineates the transition between global minima $S_0$ and $S_1$, as illustrated in Fig.~(\ref{Comparison})(j), where $S_0$ and $S_1$ attain equal energy levels $E$. Adjacent bars to the main plots list distance $\Delta d$ for journey $S_1 \rightarrow {\rm SP} \rightarrow S_0$ for VISA (blue) and CIM (red), where for CIM we take $p = \gamma$.}
    \label{Critical}
\end{figure}

To determine the critical points of $H_{\rm VISA}$, we set $\frac{\partial E}{\partial x_i} = 0$ for all $i$, where $i$ ranges from $1$ to $N$. Figure~(\ref{Critical}) depicts these critical points, with the minima corresponding to the states $S_0$ and $S_1$ emphasized. As the gain $\gamma_i (t)$ and the collinearity penalty $P(t)$ are systematically increased, a transition in the ground state from $S_1$ to $S_0$ is observed. Notably, there is an infinite continuum of $(\gamma_i, P)$ pairs where $S_0$ and $S_1$ share the same energy level. This continuum forms a demarcation line in the $\gamma - P$ space, illustrated in Fig.~(\ref{Comparison})(j). At a specific point on this boundary, as depicted in Fig.~(\ref{Critical})(c), both $S_0$ and $S_1$ emerge as  ground states. Furthermore, Fig.~(\ref{Critical}) quantifies the average Euclidean distance $\Delta d$ that each soft-spin $x_{pi}$ needs to traverse to transition from the local minimum $S_1$ to the global minimum $S_0$, passing through the nearest saddle point. In every scenario analyzed, the VISA model demonstrates a shorter requisite distance compared to the analogous CIM model, underscoring its efficiency in navigating the energy landscape.

\subsection{Basins of Attraction}

\begin{figure}[ht]
\centering
     \includegraphics[width=\columnwidth]{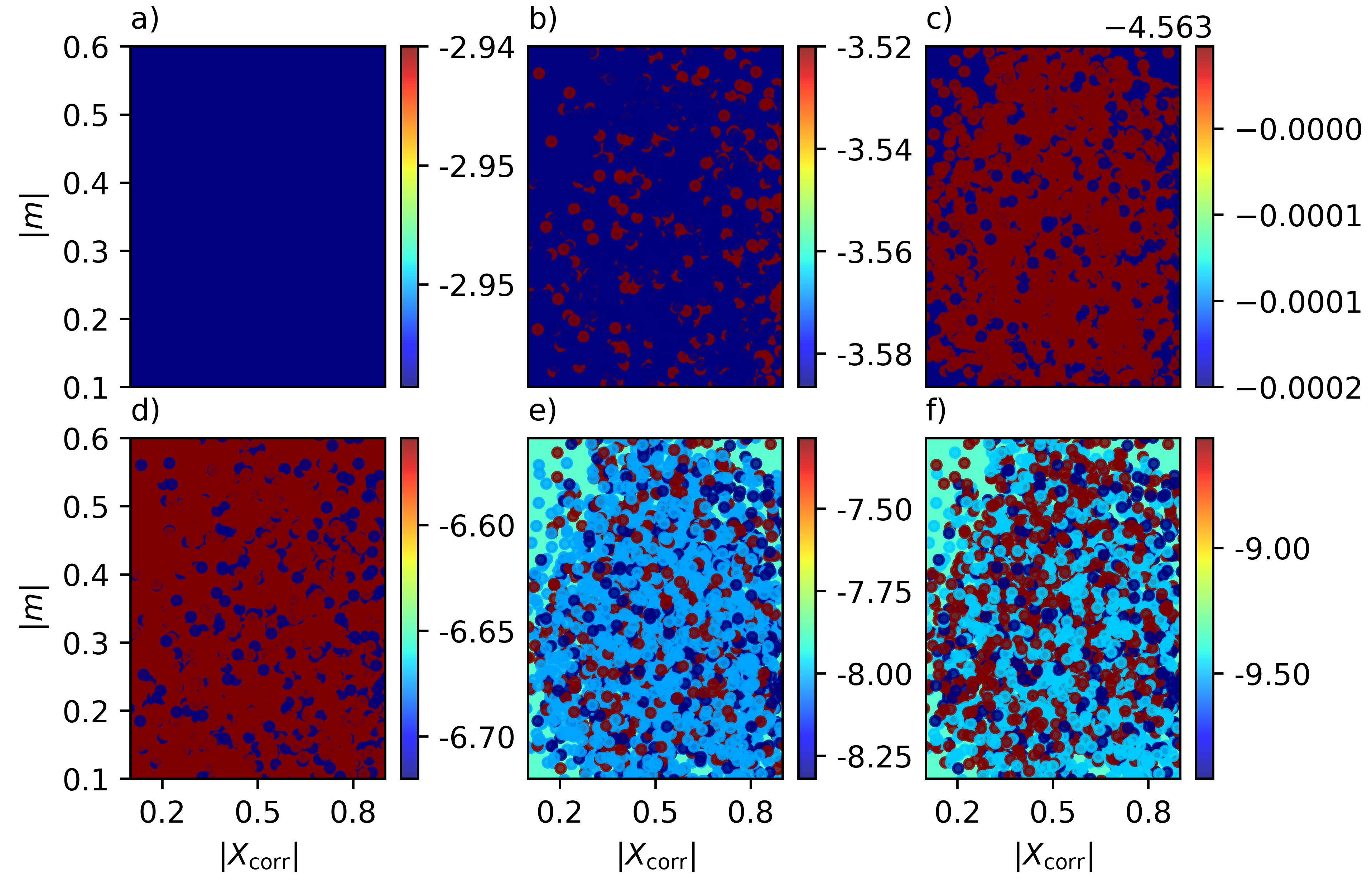}
     \caption{Basins of attraction of the VISA model landscape on $J$-M\"obius ladder graph. We take $N = 8$ and $J = 0.4$ with various gains $\gamma_i = \gamma \forall i$ and penalties $P$ as they anneal in the VISA method. Four thousand randomly distributed starting points $x_{pi} (0)$ in $[-1, 1]$ are plotted with colors corresponding to minima reached via gradient descent using Newton's method. $(\gamma, P)$ (a) $(-0.5, 0)$, (b) $(-0.25, 0.15)$, (c) $(-0.087, 0.32)$, (d) $(0.25, 0.5)$, (e) $(0.5, 0.75)$, and (f) $(0.75, 1.0)$. To characterize points, the magnetization magnitude $|{\bf m}| = \sqrt{m_{x}^{2} + m_{y}^{2} + m_{z}^{2}}$ and correlation magnitude $|{\bf X_{\rm corr}}| = \sqrt{X_{x}^{2} + X_{y}^{2} + X_{z}^{2}}$ are used, where $m_p = \sum_{i} x_{pi} / N$ and $X_{p} = \sum_{i} (x_{pi} - m_p) (x_{p(i + 1)} - m_p) / \sum_{i} (x_{pi} - m_p)^2$ are the average magnetization and correlations along the $p$-axis, respectively. For small $\gamma$ and $P$, the basin of attraction is dominated by the $S_{1}$ state. As $\gamma$ and $P$ grow, the volume of the basin of attraction for state $S_{0}$ increases. For $\gamma$ and $P$, which lie on the critical boundary, shown here in (c), both $S_{0}$ and $S_{1}$ states have the same energy. As $\gamma$ and $P$ increase further, $S_{0}$ becomes the ground state. This is indicated by the switch between excited (red) and ground (blue) states.}
    \label{Basins}
\end{figure}

Figure~(\ref{Basins}) illustrates the basins of attraction for a range of gain $\gamma_i = \gamma \forall i$ and penalty $P$ values on $J$-M\"obius ladder graphs with $J = 0.4$. Panels (a) to (f) in the figure show the evolution of these basins as $\gamma$ and $P$ are incrementally increased, following a typical gain-based and penalty annealing schedule in the VISA model. The basins of attraction are defined by the initial states $x_{pi} (0)$, uniformly distributed across the interval $[-1, 1]$, which converge to distinct minima through gradient descent. Initially, at $(\gamma, P) = (-0.5, 0)$, the global minimum is at $S_1$, which has the largest basin of attraction, capturing all initial states towards $S_{1}$. However, as $\gamma$ and $P$ progressively increase, the attraction basin associated with the excited state $S_{0}$ expands. Beyond the critical threshold in the $\gamma - P$ parameter space, demonstrated in Fig.~(\ref{Comparison})(j), $S_{0}$ transitions to become the new ground state. Further annealing of $\gamma$ and $P$ beyond this threshold reveals additional higher-energy states through the process of gradient descent.

\subsection{Second-Order Scalar Networks Minimizing the Ising Hamiltonian on $J$-M\"obius Ladder Graphs}

\begin{figure}[ht]
\centering
     \includegraphics[width=\columnwidth]{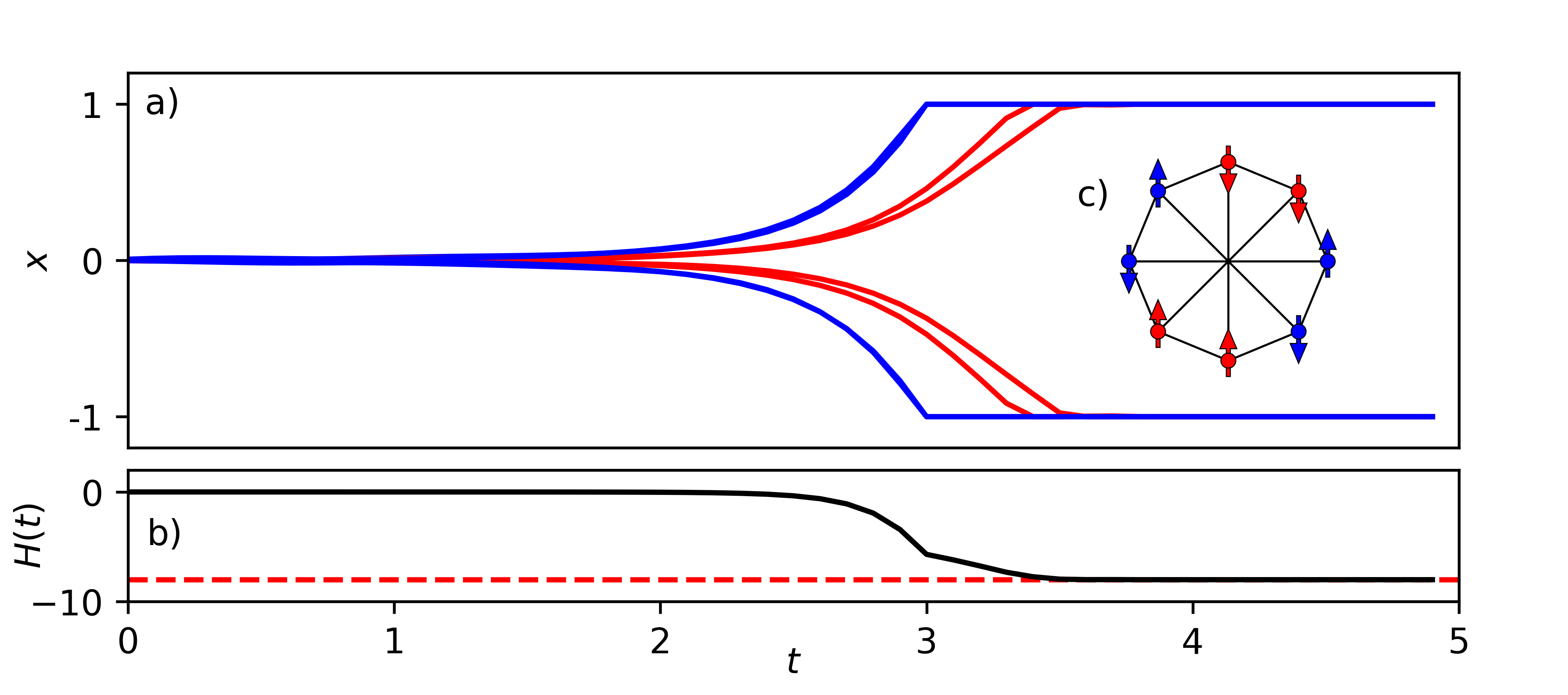}
     \caption{Evolution of momentum enhanced Hopfield-Tank method for an $N = 8$ $J$-M\"obius ladder network with $J = 1$. (a) The amplitudes connected by the frustrated edges are lower than in the rest of the system and are shown in red. (b) The corresponding Ising Hamiltonian decreases in time, and the ground state is found, with energy given by the dashed red line. (c) Schematic representation of state $S_1$ realised by the soft-spin momentum enhanced HT model. Here, $m = 1.0$, $\alpha = 2.5 / \lambda_{\rm max}$, $\beta = 1.5 (1 - t / T)$, $\gamma = 0.7$, and $x_i (0)$ is chosen uniformly at random from the range $[-0.01, 0.01]$. Fourth order Runge-Kutta is used with fixed time step $\Delta t = 0.1$ to solve Eq.~(\ref{ME-HT}).}
    \label{HT}
\end{figure}

\begin{figure}[ht]
\centering
     \includegraphics[width=\columnwidth]{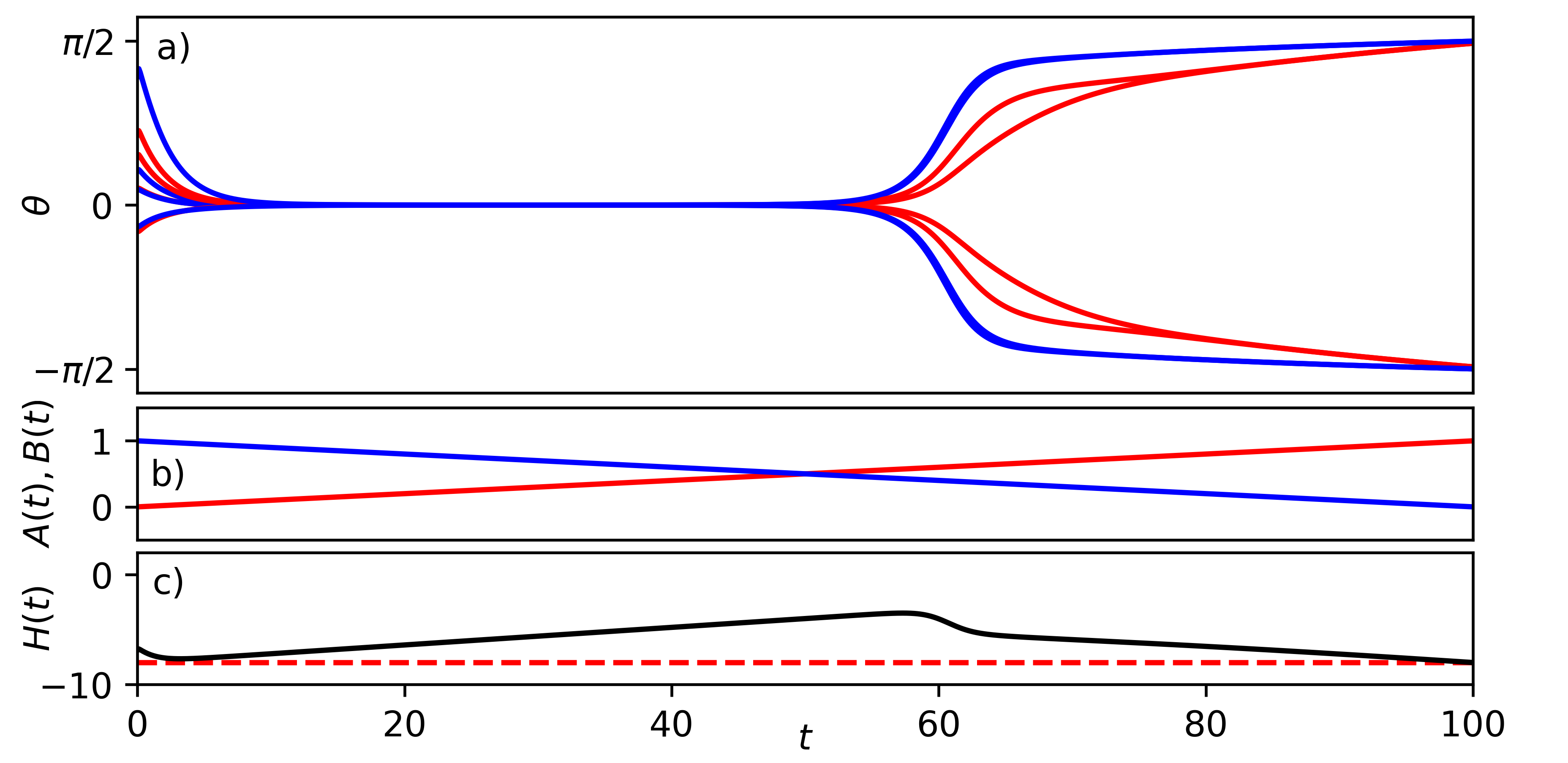}
     \caption{Evolution of the spin-vector Langevin (SVL) model. Applying the SVL model on an $N = 8$ $J$-M\"obius ladder network with coupling strength $J = 1$. (a) depicts the amplitudes connected by frustrated edges, shown in red, which are lower than those in the rest of the system. (b) illustrates the real annealing functions $A(t)$ in blue and $B(t)$ in red, satisfying the boundary conditions $A(0) = B(T) = 1$ and $A(T) = B(0) = 0$, over an annealing period $T = 100$. (c) shows the total Hamiltonian $H(t) = A(t) H_0 + B(t) H_P$ dynamics: it decreases initially as the ground state of $H_0$ is reached, then increases as $A(t)$ and $B(t)$ are annealed. Post a critical juncture, the SVL model aims to minimize the problem Hamiltonian $H_P$, identifying the ground state of the $J$-M\"obius ladder, indicated by the dashed red line. Parameters used are $m = 1.0$, $\alpha = 1 / \lambda_{\rm max}$, $\gamma = 1.0$, and noise $\xi = 0$. Annealing schedules are defined as $A(t) = 1 - t/T$ and $B(t) = t/T$. Initial phases $\theta_i(0)$ are randomly selected from $[-\pi/2, \pi/2]$.  Equation (\ref{SVL_Eq}) is solved using a fourth-order Runge-Kutta method with a fixed time step of $\Delta t = 0.1$.}
    \label{SVL}
\end{figure}

Figure~(\ref{HT}) illustrates the Aharonov-Hopf bifurcation in soft-spins within the ME-HT model. For each time step $t$, we apply a clipping function $[ {\bf x}_i ]_t = \text{sgn}([ {\bf x}_i ]_t )$ when $ | [ {\bf x}_i ]_t | > 1$, alongside a nonlinear activation function $g({\bf x}_i) = \tanh({\bf x}_i)$. The parameter $\alpha$ is adjusted according to the largest positive eigenvalue $\lambda_{\rm max}$ of the adjacency matrix ${\bf J}$. Concurrently, Fig.~(\ref{SVL}) displays the temporal evolution of scalar amplitudes as dictated by the SVL model, applied to the same M\"obius graph with $J = 1$.

\subsection{Eigenvalues, Eigenvectors, and Ground States of $J-G$ Cyclic Graphs}

\begin{figure}[ht]
\centering
     \includegraphics[width=\columnwidth]{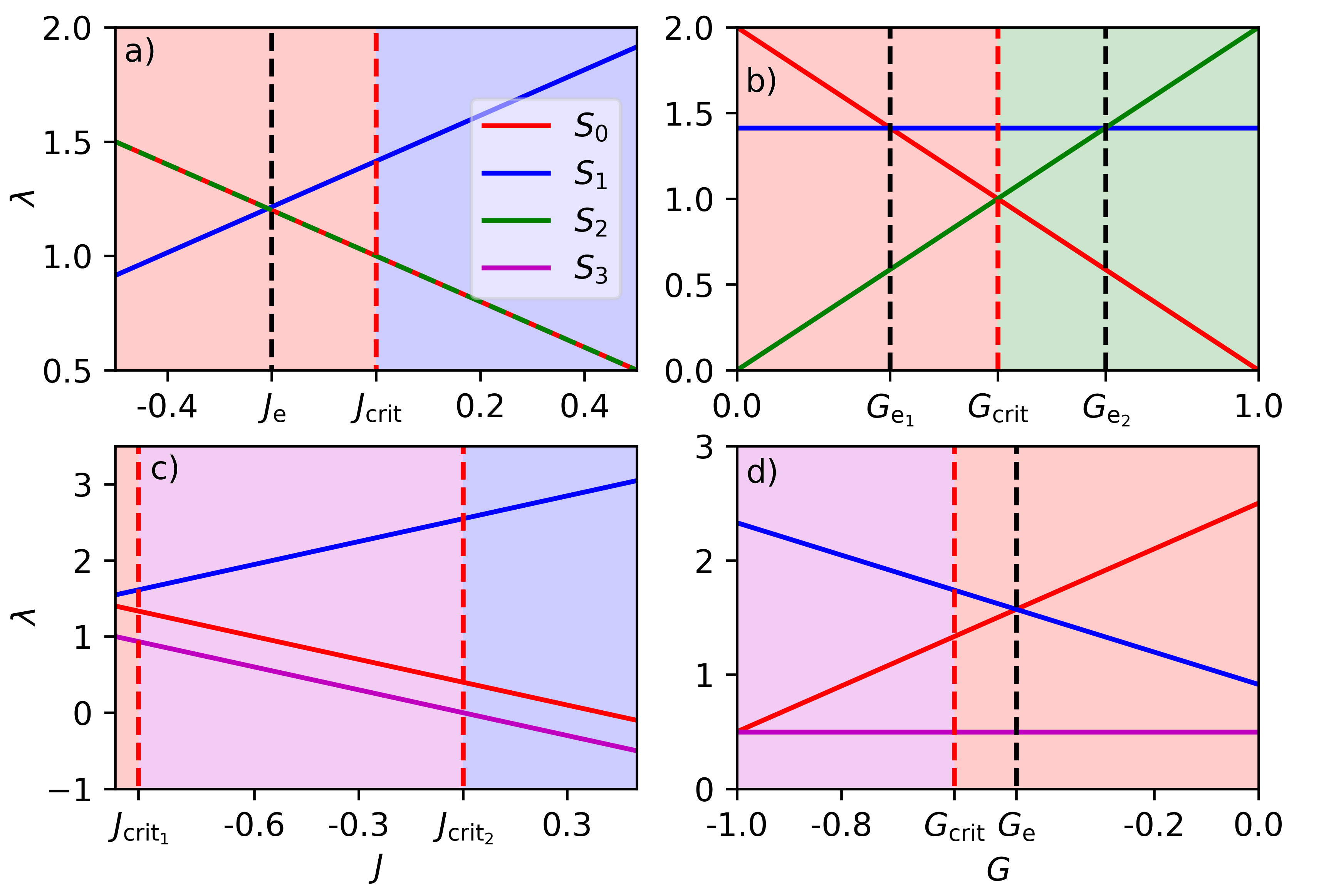}
     \caption{Eigenvalues of $N = 8$ $J-G$ cyclic graphs as functions of $J$ and $G$. Each plot corresponds to one of the red lines in Figs.~(\ref{New_Mobius})(f) and (i). Ground state probabilities are calculated on these lines in $J - G$ space for various methods in the main text. (a) $k = 2$ and $G = 0.5$. (b) $k = 2$ and $J = 0$. (c) $k = 3$ and $G = -0.8$. (d) $k = 3$ and $J = -0.5$. Vertical dashed red and black lines show where ground state energies and leading eigenvalues change, respectively. Background colors denote ground states in specific regions, with states and corresponding colors listed in the key.}
    \label{EValue_Figure}
\end{figure}

Here we refer to $J-G$ cyclic graph defined in the main text, and illustrated in Figs.~(\ref{New_Mobius})(a)-(b) for $N = 8$. For general $N$ with couplings $J_{ij} \in \{ -1,  -J, -G \}$, the $N$ eigenvalues are given by
\begin{equation} \label{EValues}
    \lambda_n = - 2 \cos (2 \pi n / N) - J (-1)^{n} - 2 G \cos (2 \pi k n / N),
\end{equation}
where $n = 1, 2, \ldots, N$. Equation~(\ref{EValues}) follows from substituting $J_{1, j}$ into the general form of matrix eigenvalues for cyclic graphs given in the main text. As $J$ and $G$ vary within $[-1, 1]$, the leading eigenvalue changes. Constrained to this domain, for $N = 8$ and $k = 2$, the leading eigenvalues are given by $n = 4, 5, 6$. Substituting these values into Eq.~(\ref{EValues}) and equating each pair gives three boundaries defined by $J = 1 - \sqrt{2}/2 - G$, $J = G - 1/\sqrt{2}$, and $G = 1/2$. Similarly for $k = 3$, we obtain two unique leading eigenvalues ($n = 4, 5$) in the $J - G$ domain, with boundary $J = 1 - \sqrt{2}/2 + (1 + 1/\sqrt{2})G$. Eigenvectors corresponding to leading eigenvalues are used to deduce analogous Ising states by projecting the eigenvector onto the nearest hypercube corner $[-1, 1]^{N}$. These Ising states are given in Figs.~(\ref{New_Mobius})(e) and (h). Gurobi is used to obtain ground states in $J - G$ space. Four unique configurations are found with $N = 8$: namely $S_0$, $S_1$, and $S_2$ for $k = 2$, and $S_0$, $S_1$, and $S_3$ for $k = 3$. $S_2$ and $S_3$ are defined as the configurations of Ising spins given by $S_2 = (1, 1, -1, -1, 1, 1, -1, -1)$ and $S_3 = (1, 1, -1, 1, 1, -1, 1, -1)$. For even $N/2$, these four states have energies
\begin{align}
    E_0 (J, G) & = (J - 2 - 2G(-1)^{k + 1}) N / 2, \\
    E_1 (J, G) & = 4 - (J + 2) N / 2 + (-1)^{k} (N - 4k) G, \\
    E_2 (J, G) & = ((-1)^{N / 4} J + (1 + (-1)^{k})(-1)^{d} G) N / 2, \\
    E_3 (J, G) & = 4 - N + (2 - N/2)J + (-1)^{k}(N - 4k)G,
\end{align}
where $d = k/2$ if $k$ is even, and $0$ otherwise. Equating the relevant energies for $k = 2$ gives three boundaries defined by separators $J = 1/2 - G$, $J = G - 1/2$, and $G = 1/2$. Similarly for $k = 3$, the boundary edges are given by $J = 1/2 + 3/2 G$, $J = 2/3 + 2G$, and $J = 0$. Eigenvalues and ground states for $J-G$ cyclic graphs are shown in Fig.~(\ref{EValue_Figure}) for values of $J$ and $G$ considered in the main text.

\bibliography{ReferencesSpacialSpins,references}

\providecommand{\noopsort}[1]{}\providecommand{\singleletter}[1]{#1}%
\begin{thebibliography}{47}%
\makeatletter
\providecommand \@ifxundefined [1]{%
 \@ifx{#1\undefined}
}%
\providecommand \@ifnum [1]{%
 \ifnum #1\expandafter \@firstoftwo
 \else \expandafter \@secondoftwo
 \fi
}%
\providecommand \@ifx [1]{%
 \ifx #1\expandafter \@firstoftwo
 \else \expandafter \@secondoftwo
 \fi
}%
\providecommand \natexlab [1]{#1}%
\providecommand \enquote  [1]{``#1''}%
\providecommand \bibnamefont  [1]{#1}%
\providecommand \bibfnamefont [1]{#1}%
\providecommand \citenamefont [1]{#1}%
\providecommand \href@noop [0]{\@secondoftwo}%
\providecommand \href [0]{\begingroup \@sanitize@url \@href}%
\providecommand \@href[1]{\@@startlink{#1}\@@href}%
\providecommand \@@href[1]{\endgroup#1\@@endlink}%
\providecommand \@sanitize@url [0]{\catcode `\\12\catcode `\$12\catcode
  `\&12\catcode `\#12\catcode `\^12\catcode `\_12\catcode `\%12\relax}%
\providecommand \@@startlink[1]{}%
\providecommand \@@endlink[0]{}%
\providecommand \url  [0]{\begingroup\@sanitize@url \@url }%
\providecommand \@url [1]{\endgroup\@href {#1}{\urlprefix }}%
\providecommand \urlprefix  [0]{URL }%
\providecommand \Eprint [0]{\href }%
\providecommand \doibase [0]{http://dx.doi.org/}%
\providecommand \selectlanguage [0]{\@gobble}%
\providecommand \bibinfo  [0]{\@secondoftwo}%
\providecommand \bibfield  [0]{\@secondoftwo}%
\providecommand \translation [1]{[#1]}%
\providecommand \BibitemOpen [0]{}%
\providecommand \bibitemStop [0]{}%
\providecommand \bibitemNoStop [0]{.\EOS\space}%
\providecommand \EOS [0]{\spacefactor3000\relax}%
\providecommand \BibitemShut  [1]{\csname bibitem#1\endcsname}%
\let\auto@bib@innerbib\@empty
\bibitem [{\citenamefont {Thompson}\ and\ \citenamefont
  {Spanuth}(2021)}]{thompson2021decline}%
  \BibitemOpen
  \bibfield  {author} {\bibinfo {author} {\bibfnamefont {N.~C.}\ \bibnamefont
  {Thompson}}\ and\ \bibinfo {author} {\bibfnamefont {S.}~\bibnamefont
  {Spanuth}},\ }\href@noop {} {\bibfield  {journal} {\bibinfo  {journal}
  {Communications of the ACM}\ }\textbf {\bibinfo {volume} {64}},\ \bibinfo
  {pages} {64} (\bibinfo {year} {2021})}\BibitemShut {NoStop}%
\bibitem [{\citenamefont {Vadlamani}\ \emph {et~al.}(2020)\citenamefont
  {Vadlamani}, \citenamefont {Xiao},\ and\ \citenamefont
  {Yablonovitch}}]{vadlamani2020physics}%
  \BibitemOpen
  \bibfield  {author} {\bibinfo {author} {\bibfnamefont {S.~K.}\ \bibnamefont
  {Vadlamani}}, \bibinfo {author} {\bibfnamefont {T.~P.}\ \bibnamefont {Xiao}},
  \ and\ \bibinfo {author} {\bibfnamefont {E.}~\bibnamefont {Yablonovitch}},\
  }\href@noop {} {\bibfield  {journal} {\bibinfo  {journal} {Proceedings of the
  National Academy of Sciences}\ }\textbf {\bibinfo {volume} {117}},\ \bibinfo
  {pages} {26639} (\bibinfo {year} {2020})}\BibitemShut {NoStop}%
\bibitem [{\citenamefont {Farhi}\ \emph {et~al.}(2000)\citenamefont {Farhi},
  \citenamefont {Goldstone}, \citenamefont {Gutmann},\ and\ \citenamefont
  {Sipser}}]{farhi}%
  \BibitemOpen
  \bibfield  {author} {\bibinfo {author} {\bibfnamefont {E.}~\bibnamefont
  {Farhi}}, \bibinfo {author} {\bibfnamefont {J.}~\bibnamefont {Goldstone}},
  \bibinfo {author} {\bibfnamefont {S.}~\bibnamefont {Gutmann}}, \ and\
  \bibinfo {author} {\bibfnamefont {M.}~\bibnamefont {Sipser}},\ }\href@noop {}
  {\bibfield  {journal} {\bibinfo  {journal} {arXiv preprint quant-ph/0001106}\
  } (\bibinfo {year} {2000})}\BibitemShut {NoStop}%
\bibitem [{\citenamefont {Lucas}(2014)}]{lucas2014ising}%
  \BibitemOpen
  \bibfield  {author} {\bibinfo {author} {\bibfnamefont {A.}~\bibnamefont
  {Lucas}},\ }\href@noop {} {\bibfield  {journal} {\bibinfo  {journal}
  {Frontiers in physics}\ }\textbf {\bibinfo {volume} {2}},\ \bibinfo {pages}
  {5} (\bibinfo {year} {2014})}\BibitemShut {NoStop}%
\bibitem [{\citenamefont {Berloff}\ \emph {et~al.}(2017)\citenamefont
  {Berloff}, \citenamefont {Silva}, \citenamefont {Kalinin}, \citenamefont
  {Askitopoulos}, \citenamefont {T{\"o}pfer}, \citenamefont {Cilibrizzi},
  \citenamefont {Langbein},\ and\ \citenamefont
  {Lagoudakis}}]{berloff2017realizing}%
  \BibitemOpen
  \bibfield  {author} {\bibinfo {author} {\bibfnamefont {N.~G.}\ \bibnamefont
  {Berloff}}, \bibinfo {author} {\bibfnamefont {M.}~\bibnamefont {Silva}},
  \bibinfo {author} {\bibfnamefont {K.}~\bibnamefont {Kalinin}}, \bibinfo
  {author} {\bibfnamefont {A.}~\bibnamefont {Askitopoulos}}, \bibinfo {author}
  {\bibfnamefont {J.~D.}\ \bibnamefont {T{\"o}pfer}}, \bibinfo {author}
  {\bibfnamefont {P.}~\bibnamefont {Cilibrizzi}}, \bibinfo {author}
  {\bibfnamefont {W.}~\bibnamefont {Langbein}}, \ and\ \bibinfo {author}
  {\bibfnamefont {P.~G.}\ \bibnamefont {Lagoudakis}},\ }\href@noop {}
  {\bibfield  {journal} {\bibinfo  {journal} {Nature materials}\ } (\bibinfo
  {year} {2017})}\BibitemShut {NoStop}%
\bibitem [{\citenamefont {Kalinin}\ \emph {et~al.}(2020)\citenamefont
  {Kalinin}, \citenamefont {Amo}, \citenamefont {Bloch},\ and\ \citenamefont
  {Berloff}}]{kalinin2020polaritonic}%
  \BibitemOpen
  \bibfield  {author} {\bibinfo {author} {\bibfnamefont {K.~P.}\ \bibnamefont
  {Kalinin}}, \bibinfo {author} {\bibfnamefont {A.}~\bibnamefont {Amo}},
  \bibinfo {author} {\bibfnamefont {J.}~\bibnamefont {Bloch}}, \ and\ \bibinfo
  {author} {\bibfnamefont {N.~G.}\ \bibnamefont {Berloff}},\ }\href@noop {}
  {\bibfield  {journal} {\bibinfo  {journal} {Nanophotonics}\ }\textbf
  {\bibinfo {volume} {9}},\ \bibinfo {pages} {4127} (\bibinfo {year}
  {2020})}\BibitemShut {NoStop}%
\bibitem [{\citenamefont {De~las Cuevas}\ and\ \citenamefont
  {Cubitt}(2016)}]{Cubitt_universality}%
  \BibitemOpen
  \bibfield  {author} {\bibinfo {author} {\bibfnamefont {G.}~\bibnamefont
  {De~las Cuevas}}\ and\ \bibinfo {author} {\bibfnamefont {T.~S.}\ \bibnamefont
  {Cubitt}},\ }\href@noop {} {\bibfield  {journal} {\bibinfo  {journal}
  {Science}\ }\textbf {\bibinfo {volume} {351}},\ \bibinfo {pages} {1180}
  (\bibinfo {year} {2016})}\BibitemShut {NoStop}%
\bibitem [{\citenamefont {McMahon}\ \emph {et~al.}(2016)\citenamefont
  {McMahon}, \citenamefont {Marandi}, \citenamefont {Haribara}, \citenamefont
  {Hamerly}, \citenamefont {Langrock}, \citenamefont {Tamate}, \citenamefont
  {Inagaki}, \citenamefont {Takesue}, \citenamefont {Utsunomiya}, \citenamefont
  {Aihara} \emph {et~al.}}]{mcmahon2016fully}%
  \BibitemOpen
  \bibfield  {author} {\bibinfo {author} {\bibfnamefont {P.~L.}\ \bibnamefont
  {McMahon}}, \bibinfo {author} {\bibfnamefont {A.}~\bibnamefont {Marandi}},
  \bibinfo {author} {\bibfnamefont {Y.}~\bibnamefont {Haribara}}, \bibinfo
  {author} {\bibfnamefont {R.}~\bibnamefont {Hamerly}}, \bibinfo {author}
  {\bibfnamefont {C.}~\bibnamefont {Langrock}}, \bibinfo {author}
  {\bibfnamefont {S.}~\bibnamefont {Tamate}}, \bibinfo {author} {\bibfnamefont
  {T.}~\bibnamefont {Inagaki}}, \bibinfo {author} {\bibfnamefont
  {H.}~\bibnamefont {Takesue}}, \bibinfo {author} {\bibfnamefont
  {S.}~\bibnamefont {Utsunomiya}}, \bibinfo {author} {\bibfnamefont
  {K.}~\bibnamefont {Aihara}},  \emph {et~al.},\ }\href@noop {} {\bibfield
  {journal} {\bibinfo  {journal} {Science}\ }\textbf {\bibinfo {volume}
  {354}},\ \bibinfo {pages} {614} (\bibinfo {year} {2016})}\BibitemShut
  {NoStop}%
\bibitem [{\citenamefont {Inagaki}\ \emph {et~al.}(2016)\citenamefont
  {Inagaki}, \citenamefont {Haribara}, \citenamefont {Igarashi}, \citenamefont
  {Sonobe}, \citenamefont {Tamate}, \citenamefont {Honjo}, \citenamefont
  {Marandi}, \citenamefont {McMahon}, \citenamefont {Umeki}, \citenamefont
  {Enbutsu} \emph {et~al.}}]{inagaki2016coherent}%
  \BibitemOpen
  \bibfield  {author} {\bibinfo {author} {\bibfnamefont {T.}~\bibnamefont
  {Inagaki}}, \bibinfo {author} {\bibfnamefont {Y.}~\bibnamefont {Haribara}},
  \bibinfo {author} {\bibfnamefont {K.}~\bibnamefont {Igarashi}}, \bibinfo
  {author} {\bibfnamefont {T.}~\bibnamefont {Sonobe}}, \bibinfo {author}
  {\bibfnamefont {S.}~\bibnamefont {Tamate}}, \bibinfo {author} {\bibfnamefont
  {T.}~\bibnamefont {Honjo}}, \bibinfo {author} {\bibfnamefont
  {A.}~\bibnamefont {Marandi}}, \bibinfo {author} {\bibfnamefont {P.~L.}\
  \bibnamefont {McMahon}}, \bibinfo {author} {\bibfnamefont {T.}~\bibnamefont
  {Umeki}}, \bibinfo {author} {\bibfnamefont {K.}~\bibnamefont {Enbutsu}},
  \emph {et~al.},\ }\href@noop {} {\bibfield  {journal} {\bibinfo  {journal}
  {Science}\ }\textbf {\bibinfo {volume} {354}},\ \bibinfo {pages} {603}
  (\bibinfo {year} {2016})}\BibitemShut {NoStop}%
\bibitem [{\citenamefont {Yamamoto}\ \emph {et~al.}(2017)\citenamefont
  {Yamamoto}, \citenamefont {Aihara}, \citenamefont {Leleu}, \citenamefont
  {Kawarabayashi}, \citenamefont {Kako}, \citenamefont {Fejer}, \citenamefont
  {Inoue},\ and\ \citenamefont {Takesue}}]{yamamoto2017coherent}%
  \BibitemOpen
  \bibfield  {author} {\bibinfo {author} {\bibfnamefont {Y.}~\bibnamefont
  {Yamamoto}}, \bibinfo {author} {\bibfnamefont {K.}~\bibnamefont {Aihara}},
  \bibinfo {author} {\bibfnamefont {T.}~\bibnamefont {Leleu}}, \bibinfo
  {author} {\bibfnamefont {K.-i.}\ \bibnamefont {Kawarabayashi}}, \bibinfo
  {author} {\bibfnamefont {S.}~\bibnamefont {Kako}}, \bibinfo {author}
  {\bibfnamefont {M.}~\bibnamefont {Fejer}}, \bibinfo {author} {\bibfnamefont
  {K.}~\bibnamefont {Inoue}}, \ and\ \bibinfo {author} {\bibfnamefont
  {H.}~\bibnamefont {Takesue}},\ }\href@noop {} {\bibfield  {journal} {\bibinfo
   {journal} {npj Quantum Information}\ }\textbf {\bibinfo {volume} {3}},\
  \bibinfo {pages} {1} (\bibinfo {year} {2017})}\BibitemShut {NoStop}%
\bibitem [{\citenamefont {Honjo}\ \emph {et~al.}(2021)\citenamefont {Honjo},
  \citenamefont {Sonobe}, \citenamefont {Inaba}, \citenamefont {Inagaki},
  \citenamefont {Ikuta}, \citenamefont {Yamada}, \citenamefont {Kazama},
  \citenamefont {Enbutsu}, \citenamefont {Umeki}, \citenamefont {Kasahara}
  \emph {et~al.}}]{honjo2021100}%
  \BibitemOpen
  \bibfield  {author} {\bibinfo {author} {\bibfnamefont {T.}~\bibnamefont
  {Honjo}}, \bibinfo {author} {\bibfnamefont {T.}~\bibnamefont {Sonobe}},
  \bibinfo {author} {\bibfnamefont {K.}~\bibnamefont {Inaba}}, \bibinfo
  {author} {\bibfnamefont {T.}~\bibnamefont {Inagaki}}, \bibinfo {author}
  {\bibfnamefont {T.}~\bibnamefont {Ikuta}}, \bibinfo {author} {\bibfnamefont
  {Y.}~\bibnamefont {Yamada}}, \bibinfo {author} {\bibfnamefont
  {T.}~\bibnamefont {Kazama}}, \bibinfo {author} {\bibfnamefont
  {K.}~\bibnamefont {Enbutsu}}, \bibinfo {author} {\bibfnamefont
  {T.}~\bibnamefont {Umeki}}, \bibinfo {author} {\bibfnamefont
  {R.}~\bibnamefont {Kasahara}},  \emph {et~al.},\ }\href@noop {} {\bibfield
  {journal} {\bibinfo  {journal} {Science advances}\ }\textbf {\bibinfo
  {volume} {7}},\ \bibinfo {pages} {eabh0952} (\bibinfo {year}
  {2021})}\BibitemShut {NoStop}%
\bibitem [{\citenamefont {Cai}\ \emph {et~al.}(2020)\citenamefont {Cai},
  \citenamefont {Kumar}, \citenamefont {Van~Vaerenbergh}, \citenamefont
  {Sheng}, \citenamefont {Liu}, \citenamefont {Li}, \citenamefont {Liu},
  \citenamefont {Foltin}, \citenamefont {Yu}, \citenamefont {Xia} \emph
  {et~al.}}]{cai2020power}%
  \BibitemOpen
  \bibfield  {author} {\bibinfo {author} {\bibfnamefont {F.}~\bibnamefont
  {Cai}}, \bibinfo {author} {\bibfnamefont {S.}~\bibnamefont {Kumar}}, \bibinfo
  {author} {\bibfnamefont {T.}~\bibnamefont {Van~Vaerenbergh}}, \bibinfo
  {author} {\bibfnamefont {X.}~\bibnamefont {Sheng}}, \bibinfo {author}
  {\bibfnamefont {R.}~\bibnamefont {Liu}}, \bibinfo {author} {\bibfnamefont
  {C.}~\bibnamefont {Li}}, \bibinfo {author} {\bibfnamefont {Z.}~\bibnamefont
  {Liu}}, \bibinfo {author} {\bibfnamefont {M.}~\bibnamefont {Foltin}},
  \bibinfo {author} {\bibfnamefont {S.}~\bibnamefont {Yu}}, \bibinfo {author}
  {\bibfnamefont {Q.}~\bibnamefont {Xia}},  \emph {et~al.},\ }\href@noop {}
  {\bibfield  {journal} {\bibinfo  {journal} {Nature Electronics}\ }\textbf
  {\bibinfo {volume} {3}},\ \bibinfo {pages} {409} (\bibinfo {year}
  {2020})}\BibitemShut {NoStop}%
\bibitem [{\citenamefont {Babaeian}\ \emph {et~al.}(2019)\citenamefont
  {Babaeian}, \citenamefont {Nguyen}, \citenamefont {Demir}, \citenamefont
  {Akbulut}, \citenamefont {Blanche}, \citenamefont {Kaneda}, \citenamefont
  {Guha}, \citenamefont {Neifeld},\ and\ \citenamefont
  {Peyghambarian}}]{babaeian2019single}%
  \BibitemOpen
  \bibfield  {author} {\bibinfo {author} {\bibfnamefont {M.}~\bibnamefont
  {Babaeian}}, \bibinfo {author} {\bibfnamefont {D.~T.}\ \bibnamefont
  {Nguyen}}, \bibinfo {author} {\bibfnamefont {V.}~\bibnamefont {Demir}},
  \bibinfo {author} {\bibfnamefont {M.}~\bibnamefont {Akbulut}}, \bibinfo
  {author} {\bibfnamefont {P.-A.}\ \bibnamefont {Blanche}}, \bibinfo {author}
  {\bibfnamefont {Y.}~\bibnamefont {Kaneda}}, \bibinfo {author} {\bibfnamefont
  {S.}~\bibnamefont {Guha}}, \bibinfo {author} {\bibfnamefont {M.~A.}\
  \bibnamefont {Neifeld}}, \ and\ \bibinfo {author} {\bibfnamefont
  {N.}~\bibnamefont {Peyghambarian}},\ }\href@noop {} {\bibfield  {journal}
  {\bibinfo  {journal} {Nature communications}\ }\textbf {\bibinfo {volume}
  {10}},\ \bibinfo {pages} {1} (\bibinfo {year} {2019})}\BibitemShut {NoStop}%
\bibitem [{\citenamefont {Pal}\ \emph {et~al.}(2020)\citenamefont {Pal},
  \citenamefont {Mahler}, \citenamefont {Tradonsky}, \citenamefont {Friesem},\
  and\ \citenamefont {Davidson}}]{pal2020rapid}%
  \BibitemOpen
  \bibfield  {author} {\bibinfo {author} {\bibfnamefont {V.}~\bibnamefont
  {Pal}}, \bibinfo {author} {\bibfnamefont {S.}~\bibnamefont {Mahler}},
  \bibinfo {author} {\bibfnamefont {C.}~\bibnamefont {Tradonsky}}, \bibinfo
  {author} {\bibfnamefont {A.~A.}\ \bibnamefont {Friesem}}, \ and\ \bibinfo
  {author} {\bibfnamefont {N.}~\bibnamefont {Davidson}},\ }\href@noop {}
  {\bibfield  {journal} {\bibinfo  {journal} {Physical Review Research}\
  }\textbf {\bibinfo {volume} {2}},\ \bibinfo {pages} {033008} (\bibinfo {year}
  {2020})}\BibitemShut {NoStop}%
\bibitem [{\citenamefont {Parto}\ \emph {et~al.}(2020)\citenamefont {Parto},
  \citenamefont {Hayenga}, \citenamefont {Marandi}, \citenamefont
  {Christodoulides},\ and\ \citenamefont {Khajavikhan}}]{parto2020realizing}%
  \BibitemOpen
  \bibfield  {author} {\bibinfo {author} {\bibfnamefont {M.}~\bibnamefont
  {Parto}}, \bibinfo {author} {\bibfnamefont {W.}~\bibnamefont {Hayenga}},
  \bibinfo {author} {\bibfnamefont {A.}~\bibnamefont {Marandi}}, \bibinfo
  {author} {\bibfnamefont {D.~N.}\ \bibnamefont {Christodoulides}}, \ and\
  \bibinfo {author} {\bibfnamefont {M.}~\bibnamefont {Khajavikhan}},\
  }\href@noop {} {\bibfield  {journal} {\bibinfo  {journal} {Nature materials}\
  }\textbf {\bibinfo {volume} {19}},\ \bibinfo {pages} {725} (\bibinfo {year}
  {2020})}\BibitemShut {NoStop}%
\bibitem [{\citenamefont {Pierangeli}\ \emph {et~al.}(2019)\citenamefont
  {Pierangeli}, \citenamefont {Marcucci},\ and\ \citenamefont
  {Conti}}]{pierangeli2019large}%
  \BibitemOpen
  \bibfield  {author} {\bibinfo {author} {\bibfnamefont {D.}~\bibnamefont
  {Pierangeli}}, \bibinfo {author} {\bibfnamefont {G.}~\bibnamefont
  {Marcucci}}, \ and\ \bibinfo {author} {\bibfnamefont {C.}~\bibnamefont
  {Conti}},\ }\href@noop {} {\bibfield  {journal} {\bibinfo  {journal}
  {Physical review letters}\ }\textbf {\bibinfo {volume} {122}},\ \bibinfo
  {pages} {213902} (\bibinfo {year} {2019})}\BibitemShut {NoStop}%
\bibitem [{\citenamefont {Roques-Carmes}\ \emph {et~al.}(2020)\citenamefont
  {Roques-Carmes}, \citenamefont {Shen}, \citenamefont {Zanoci}, \citenamefont
  {Prabhu}, \citenamefont {Atieh}, \citenamefont {Jing}, \citenamefont
  {Dub{\v{c}}ek}, \citenamefont {Mao}, \citenamefont {Johnson}, \citenamefont
  {{\v{C}}eperi{\'c}} \emph {et~al.}}]{roques2020heuristic}%
  \BibitemOpen
  \bibfield  {author} {\bibinfo {author} {\bibfnamefont {C.}~\bibnamefont
  {Roques-Carmes}}, \bibinfo {author} {\bibfnamefont {Y.}~\bibnamefont {Shen}},
  \bibinfo {author} {\bibfnamefont {C.}~\bibnamefont {Zanoci}}, \bibinfo
  {author} {\bibfnamefont {M.}~\bibnamefont {Prabhu}}, \bibinfo {author}
  {\bibfnamefont {F.}~\bibnamefont {Atieh}}, \bibinfo {author} {\bibfnamefont
  {L.}~\bibnamefont {Jing}}, \bibinfo {author} {\bibfnamefont {T.}~\bibnamefont
  {Dub{\v{c}}ek}}, \bibinfo {author} {\bibfnamefont {C.}~\bibnamefont {Mao}},
  \bibinfo {author} {\bibfnamefont {M.~R.}\ \bibnamefont {Johnson}}, \bibinfo
  {author} {\bibfnamefont {V.}~\bibnamefont {{\v{C}}eperi{\'c}}},  \emph
  {et~al.},\ }\href@noop {} {\bibfield  {journal} {\bibinfo  {journal} {Nature
  communications}\ }\textbf {\bibinfo {volume} {11}},\ \bibinfo {pages} {249}
  (\bibinfo {year} {2020})}\BibitemShut {NoStop}%
\bibitem [{\citenamefont {Vretenar}\ \emph {et~al.}(2021)\citenamefont
  {Vretenar}, \citenamefont {Kassenberg}, \citenamefont {Bissesar},
  \citenamefont {Toebes},\ and\ \citenamefont
  {Klaers}}]{vretenar2021controllable}%
  \BibitemOpen
  \bibfield  {author} {\bibinfo {author} {\bibfnamefont {M.}~\bibnamefont
  {Vretenar}}, \bibinfo {author} {\bibfnamefont {B.}~\bibnamefont
  {Kassenberg}}, \bibinfo {author} {\bibfnamefont {S.}~\bibnamefont
  {Bissesar}}, \bibinfo {author} {\bibfnamefont {C.}~\bibnamefont {Toebes}}, \
  and\ \bibinfo {author} {\bibfnamefont {J.}~\bibnamefont {Klaers}},\
  }\href@noop {} {\bibfield  {journal} {\bibinfo  {journal} {Physical Review
  Research}\ }\textbf {\bibinfo {volume} {3}},\ \bibinfo {pages} {023167}
  (\bibinfo {year} {2021})}\BibitemShut {NoStop}%
\bibitem [{\citenamefont {Kalinin}\ \emph {et~al.}(2023)\citenamefont
  {Kalinin}, \citenamefont {Mourgias-Alexandris}, \citenamefont {Ballani},
  \citenamefont {Berloff}, \citenamefont {Clegg}, \citenamefont {Cletheroe},
  \citenamefont {Gkantsidis}, \citenamefont {Haller}, \citenamefont
  {Lyutsarev}, \citenamefont {Parmigiani}, \citenamefont {Pickup} \emph
  {et~al.}}]{kalinin2023analog}%
  \BibitemOpen
  \bibfield  {author} {\bibinfo {author} {\bibfnamefont {K.}~\bibnamefont
  {Kalinin}}, \bibinfo {author} {\bibfnamefont {G.}~\bibnamefont
  {Mourgias-Alexandris}}, \bibinfo {author} {\bibfnamefont {H.}~\bibnamefont
  {Ballani}}, \bibinfo {author} {\bibfnamefont {N.~G.}\ \bibnamefont
  {Berloff}}, \bibinfo {author} {\bibfnamefont {J.~H.}\ \bibnamefont {Clegg}},
  \bibinfo {author} {\bibfnamefont {D.}~\bibnamefont {Cletheroe}}, \bibinfo
  {author} {\bibfnamefont {C.}~\bibnamefont {Gkantsidis}}, \bibinfo {author}
  {\bibfnamefont {I.}~\bibnamefont {Haller}}, \bibinfo {author} {\bibfnamefont
  {V.}~\bibnamefont {Lyutsarev}}, \bibinfo {author} {\bibfnamefont
  {F.}~\bibnamefont {Parmigiani}}, \bibinfo {author} {\bibfnamefont
  {L.}~\bibnamefont {Pickup}},  \emph {et~al.},\ }\href@noop {} {\bibfield
  {journal} {\bibinfo  {journal} {arXiv preprint arXiv:2304.12594}\ } (\bibinfo
  {year} {2023})}\BibitemShut {NoStop}%
\bibitem [{\citenamefont {Goto}\ \emph {et~al.}(2021)\citenamefont {Goto},
  \citenamefont {Endo}, \citenamefont {Suzuki}, \citenamefont {Sakai},
  \citenamefont {Kanao}, \citenamefont {Hamakawa}, \citenamefont {Hidaka},
  \citenamefont {Yamasaki},\ and\ \citenamefont {Tatsumura}}]{goto2021high}%
  \BibitemOpen
  \bibfield  {author} {\bibinfo {author} {\bibfnamefont {H.}~\bibnamefont
  {Goto}}, \bibinfo {author} {\bibfnamefont {K.}~\bibnamefont {Endo}}, \bibinfo
  {author} {\bibfnamefont {M.}~\bibnamefont {Suzuki}}, \bibinfo {author}
  {\bibfnamefont {Y.}~\bibnamefont {Sakai}}, \bibinfo {author} {\bibfnamefont
  {T.}~\bibnamefont {Kanao}}, \bibinfo {author} {\bibfnamefont
  {Y.}~\bibnamefont {Hamakawa}}, \bibinfo {author} {\bibfnamefont
  {R.}~\bibnamefont {Hidaka}}, \bibinfo {author} {\bibfnamefont
  {M.}~\bibnamefont {Yamasaki}}, \ and\ \bibinfo {author} {\bibfnamefont
  {K.}~\bibnamefont {Tatsumura}},\ }\href@noop {} {\bibfield  {journal}
  {\bibinfo  {journal} {Science Advances}\ }\textbf {\bibinfo {volume} {7}},\
  \bibinfo {pages} {eabe7953} (\bibinfo {year} {2021})}\BibitemShut {NoStop}%
\bibitem [{\citenamefont {Date}\ \emph {et~al.}(2021)\citenamefont {Date},
  \citenamefont {Arthur},\ and\ \citenamefont {Pusey-Nazzaro}}]{date2021qubo}%
  \BibitemOpen
  \bibfield  {author} {\bibinfo {author} {\bibfnamefont {P.}~\bibnamefont
  {Date}}, \bibinfo {author} {\bibfnamefont {D.}~\bibnamefont {Arthur}}, \ and\
  \bibinfo {author} {\bibfnamefont {L.}~\bibnamefont {Pusey-Nazzaro}},\
  }\href@noop {} {\bibfield  {journal} {\bibinfo  {journal} {Scientific
  reports}\ }\textbf {\bibinfo {volume} {11}},\ \bibinfo {pages} {10029}
  (\bibinfo {year} {2021})}\BibitemShut {NoStop}%
\bibitem [{\citenamefont {Gilli}\ \emph {et~al.}(2019)\citenamefont {Gilli},
  \citenamefont {Maringer},\ and\ \citenamefont
  {Schumann}}]{gilli2019numerical}%
  \BibitemOpen
  \bibfield  {author} {\bibinfo {author} {\bibfnamefont {M.}~\bibnamefont
  {Gilli}}, \bibinfo {author} {\bibfnamefont {D.}~\bibnamefont {Maringer}}, \
  and\ \bibinfo {author} {\bibfnamefont {E.}~\bibnamefont {Schumann}},\
  }\href@noop {} {\emph {\bibinfo {title} {Numerical methods and optimization
  in finance}}}\ (\bibinfo  {publisher} {Academic Press},\ \bibinfo {year}
  {2019})\BibitemShut {NoStop}%
\bibitem [{\citenamefont {Pierce}\ and\ \citenamefont
  {Winfree}(2002)}]{pierce2002protein}%
  \BibitemOpen
  \bibfield  {author} {\bibinfo {author} {\bibfnamefont {N.~A.}\ \bibnamefont
  {Pierce}}\ and\ \bibinfo {author} {\bibfnamefont {E.}~\bibnamefont
  {Winfree}},\ }\href@noop {} {\bibfield  {journal} {\bibinfo  {journal}
  {Protein engineering}\ }\textbf {\bibinfo {volume} {15}},\ \bibinfo {pages}
  {779} (\bibinfo {year} {2002})}\BibitemShut {NoStop}%
\bibitem [{\citenamefont {Dill}\ \emph {et~al.}(2008)\citenamefont {Dill},
  \citenamefont {Ozkan}, \citenamefont {Shell},\ and\ \citenamefont
  {Weikl}}]{dill2008protein}%
  \BibitemOpen
  \bibfield  {author} {\bibinfo {author} {\bibfnamefont {K.~A.}\ \bibnamefont
  {Dill}}, \bibinfo {author} {\bibfnamefont {S.~B.}\ \bibnamefont {Ozkan}},
  \bibinfo {author} {\bibfnamefont {M.~S.}\ \bibnamefont {Shell}}, \ and\
  \bibinfo {author} {\bibfnamefont {T.~R.}\ \bibnamefont {Weikl}},\ }\href@noop
  {} {\bibfield  {journal} {\bibinfo  {journal} {Annu. Rev. Biophys.}\ }\textbf
  {\bibinfo {volume} {37}},\ \bibinfo {pages} {289} (\bibinfo {year}
  {2008})}\BibitemShut {NoStop}%
\bibitem [{\citenamefont {Calvanese~Strinati}\ and\ \citenamefont
  {Conti}(2022)}]{calvanese2022multidimensional}%
  \BibitemOpen
  \bibfield  {author} {\bibinfo {author} {\bibfnamefont {M.}~\bibnamefont
  {Calvanese~Strinati}}\ and\ \bibinfo {author} {\bibfnamefont
  {C.}~\bibnamefont {Conti}},\ }\href@noop {} {\bibfield  {journal} {\bibinfo
  {journal} {Nature Communications}\ }\textbf {\bibinfo {volume} {13}},\
  \bibinfo {pages} {7248} (\bibinfo {year} {2022})}\BibitemShut {NoStop}%
\bibitem [{\citenamefont {Strinati}\ and\ \citenamefont
  {Conti}(2024)}]{strinati2024hyperscaling}%
  \BibitemOpen
  \bibfield  {author} {\bibinfo {author} {\bibfnamefont {M.~C.}\ \bibnamefont
  {Strinati}}\ and\ \bibinfo {author} {\bibfnamefont {C.}~\bibnamefont
  {Conti}},\ }\href@noop {} {\bibfield  {journal} {\bibinfo  {journal}
  {Physical Review Letters}\ }\textbf {\bibinfo {volume} {132}},\ \bibinfo
  {pages} {017301} (\bibinfo {year} {2024})}\BibitemShut {NoStop}%
\bibitem [{\citenamefont {Kalinin}\ and\ \citenamefont
  {Berloff}(2018)}]{kalinin2018networks}%
  \BibitemOpen
  \bibfield  {author} {\bibinfo {author} {\bibfnamefont {K.~P.}\ \bibnamefont
  {Kalinin}}\ and\ \bibinfo {author} {\bibfnamefont {N.~G.}\ \bibnamefont
  {Berloff}},\ }\href@noop {} {\bibfield  {journal} {\bibinfo  {journal} {New
  Journal of Physics}\ }\textbf {\bibinfo {volume} {20}},\ \bibinfo {pages}
  {113023} (\bibinfo {year} {2018})}\BibitemShut {NoStop}%
\bibitem [{\citenamefont {Stroev}\ and\ \citenamefont
  {Berloff}(2021)}]{stroev2021discrete}%
  \BibitemOpen
  \bibfield  {author} {\bibinfo {author} {\bibfnamefont {N.}~\bibnamefont
  {Stroev}}\ and\ \bibinfo {author} {\bibfnamefont {N.~G.}\ \bibnamefont
  {Berloff}},\ }\href@noop {} {\bibfield  {journal} {\bibinfo  {journal}
  {Physical Review Letters}\ }\textbf {\bibinfo {volume} {126}},\ \bibinfo
  {pages} {050504} (\bibinfo {year} {2021})}\BibitemShut {NoStop}%
\bibitem [{\citenamefont {Rosenberg}(1975)}]{rosenberg1975reduction}%
  \BibitemOpen
  \bibfield  {author} {\bibinfo {author} {\bibfnamefont {I.~G.}\ \bibnamefont
  {Rosenberg}},\ }\href@noop {} {\  (\bibinfo {year} {1975})}\BibitemShut
  {NoStop}%
\bibitem [{\citenamefont {Boros}\ and\ \citenamefont
  {Hammer}(2002)}]{boros2002pseudo}%
  \BibitemOpen
  \bibfield  {author} {\bibinfo {author} {\bibfnamefont {E.}~\bibnamefont
  {Boros}}\ and\ \bibinfo {author} {\bibfnamefont {P.~L.}\ \bibnamefont
  {Hammer}},\ }\href@noop {} {\bibfield  {journal} {\bibinfo  {journal}
  {Discrete applied mathematics}\ }\textbf {\bibinfo {volume} {123}},\ \bibinfo
  {pages} {155} (\bibinfo {year} {2002})}\BibitemShut {NoStop}%
\bibitem [{\citenamefont {Hopfield}(1982)}]{hopfield1982neural}%
  \BibitemOpen
  \bibfield  {author} {\bibinfo {author} {\bibfnamefont {J.~J.}\ \bibnamefont
  {Hopfield}},\ }\href@noop {} {\bibfield  {journal} {\bibinfo  {journal}
  {Proceedings of the national academy of sciences}\ }\textbf {\bibinfo
  {volume} {79}},\ \bibinfo {pages} {2554} (\bibinfo {year}
  {1982})}\BibitemShut {NoStop}%
\bibitem [{\citenamefont {Hopfield}\ and\ \citenamefont
  {Tank}(1985)}]{hopfield1985neural}%
  \BibitemOpen
  \bibfield  {author} {\bibinfo {author} {\bibfnamefont {J.~J.}\ \bibnamefont
  {Hopfield}}\ and\ \bibinfo {author} {\bibfnamefont {D.~W.}\ \bibnamefont
  {Tank}},\ }\href@noop {} {\bibfield  {journal} {\bibinfo  {journal}
  {Biological cybernetics}\ }\textbf {\bibinfo {volume} {52}},\ \bibinfo
  {pages} {141} (\bibinfo {year} {1985})}\BibitemShut {NoStop}%
\bibitem [{\citenamefont {Goto}(2016)}]{goto2016bifurcation}%
  \BibitemOpen
  \bibfield  {author} {\bibinfo {author} {\bibfnamefont {H.}~\bibnamefont
  {Goto}},\ }\href@noop {} {\bibfield  {journal} {\bibinfo  {journal}
  {Scientific reports}\ }\textbf {\bibinfo {volume} {6}},\ \bibinfo {pages} {1}
  (\bibinfo {year} {2016})}\BibitemShut {NoStop}%
\bibitem [{\citenamefont {Cummins}\ \emph {et~al.}(2023)\citenamefont
  {Cummins}, \citenamefont {Salman},\ and\ \citenamefont
  {Berloff}}]{cummins2023classical}%
  \BibitemOpen
  \bibfield  {author} {\bibinfo {author} {\bibfnamefont {J.~S.}\ \bibnamefont
  {Cummins}}, \bibinfo {author} {\bibfnamefont {H.}~\bibnamefont {Salman}}, \
  and\ \bibinfo {author} {\bibfnamefont {N.~G.}\ \bibnamefont {Berloff}},\
  }\href@noop {} {\bibfield  {journal} {\bibinfo  {journal} {arXiv preprint
  arXiv:2311.17359}\ } (\bibinfo {year} {2023})}\BibitemShut {NoStop}%
\bibitem [{\citenamefont {Kalinin}\ and\ \citenamefont
  {Berloff}(2022)}]{kalinin2020complexity}%
  \BibitemOpen
  \bibfield  {author} {\bibinfo {author} {\bibfnamefont {K.~P.}\ \bibnamefont
  {Kalinin}}\ and\ \bibinfo {author} {\bibfnamefont {N.~G.}\ \bibnamefont
  {Berloff}},\ }\href@noop {} {\bibfield  {journal} {\bibinfo  {journal}
  {Communications Physics}\ }\textbf {\bibinfo {volume} {5}},\ \bibinfo {pages}
  {20} (\bibinfo {year} {2022})}\BibitemShut {NoStop}%
\bibitem [{\citenamefont {Orvieto}\ and\ \citenamefont
  {Lucchi}(2019)}]{orvieto2019shadowing}%
  \BibitemOpen
  \bibfield  {author} {\bibinfo {author} {\bibfnamefont {A.}~\bibnamefont
  {Orvieto}}\ and\ \bibinfo {author} {\bibfnamefont {A.}~\bibnamefont
  {Lucchi}},\ }\href@noop {} {\bibfield  {journal} {\bibinfo  {journal}
  {Advances in Neural Information Processing Systems}\ }\textbf {\bibinfo
  {volume} {32}} (\bibinfo {year} {2019})}\BibitemShut {NoStop}%
\bibitem [{\citenamefont {Saab~Jr}\ \emph {et~al.}(2022)\citenamefont
  {Saab~Jr}, \citenamefont {Phoha}, \citenamefont {Zhu},\ and\ \citenamefont
  {Ray}}]{saab2022adaptive}%
  \BibitemOpen
  \bibfield  {author} {\bibinfo {author} {\bibfnamefont {S.}~\bibnamefont
  {Saab~Jr}}, \bibinfo {author} {\bibfnamefont {S.}~\bibnamefont {Phoha}},
  \bibinfo {author} {\bibfnamefont {M.}~\bibnamefont {Zhu}}, \ and\ \bibinfo
  {author} {\bibfnamefont {A.}~\bibnamefont {Ray}},\ }\href@noop {} {\bibfield
  {journal} {\bibinfo  {journal} {Machine Learning}\ }\textbf {\bibinfo
  {volume} {111}},\ \bibinfo {pages} {3245} (\bibinfo {year}
  {2022})}\BibitemShut {NoStop}%
\bibitem [{\citenamefont {Subires}\ \emph {et~al.}(2022)\citenamefont
  {Subires}, \citenamefont {G{\'o}mez-Ruiz}, \citenamefont {Ruiz-Garc{\'\i}a},
  \citenamefont {Alonso},\ and\ \citenamefont
  {Del~Campo}}]{subires2022benchmarking}%
  \BibitemOpen
  \bibfield  {author} {\bibinfo {author} {\bibfnamefont {D.}~\bibnamefont
  {Subires}}, \bibinfo {author} {\bibfnamefont {F.~J.}\ \bibnamefont
  {G{\'o}mez-Ruiz}}, \bibinfo {author} {\bibfnamefont {A.}~\bibnamefont
  {Ruiz-Garc{\'\i}a}}, \bibinfo {author} {\bibfnamefont {D.}~\bibnamefont
  {Alonso}}, \ and\ \bibinfo {author} {\bibfnamefont {A.}~\bibnamefont
  {Del~Campo}},\ }\href@noop {} {\bibfield  {journal} {\bibinfo  {journal}
  {Physical Review Research}\ }\textbf {\bibinfo {volume} {4}},\ \bibinfo
  {pages} {023104} (\bibinfo {year} {2022})}\BibitemShut {NoStop}%
\bibitem [{\citenamefont {Jiang}\ and\ \citenamefont
  {Chu}(2022)}]{jiang2022solving}%
  \BibitemOpen
  \bibfield  {author} {\bibinfo {author} {\bibfnamefont {J.-R.}\ \bibnamefont
  {Jiang}}\ and\ \bibinfo {author} {\bibfnamefont {C.-W.}\ \bibnamefont
  {Chu}},\ }in\ \href@noop {} {\emph {\bibinfo {booktitle} {2022 IEEE 4th
  ECICE}}}\ (\bibinfo {organization} {IEEE},\ \bibinfo {year} {2022})\ pp.\
  \bibinfo {pages} {406--411}\BibitemShut {NoStop}%
\bibitem [{bfg()}]{bfgsnote}%
  \BibitemOpen
  \href@noop {} {}\bibinfo {note} {The BFGS algorithm is implemented with
  minimizer \texttt{FindMinimum} in Wolfram Mathematica. VISA Hamiltonian
  $H_{\rm VISA}$ is subject to BFGS with fixed $\gamma_i = \gamma = 1.0 \forall
  i$ and $P = 0.8$.}\BibitemShut {Stop}%
\bibitem [{\citenamefont {Gancio}\ and\ \citenamefont
  {Rubido}(2022)}]{gancio2022critical}%
  \BibitemOpen
  \bibfield  {author} {\bibinfo {author} {\bibfnamefont {J.}~\bibnamefont
  {Gancio}}\ and\ \bibinfo {author} {\bibfnamefont {N.}~\bibnamefont
  {Rubido}},\ }\href@noop {} {\bibfield  {journal} {\bibinfo  {journal} {Chaos,
  Solitons \& Fractals}\ }\textbf {\bibinfo {volume} {158}},\ \bibinfo {pages}
  {112001} (\bibinfo {year} {2022})}\BibitemShut {NoStop}%
\bibitem [{\citenamefont {Haribara}\ \emph {et~al.}(2017)\citenamefont
  {Haribara}, \citenamefont {Ishikawa}, \citenamefont {Utsunomiya},
  \citenamefont {Aihara},\ and\ \citenamefont
  {Yamamoto}}]{haribara2017performance}%
  \BibitemOpen
  \bibfield  {author} {\bibinfo {author} {\bibfnamefont {Y.}~\bibnamefont
  {Haribara}}, \bibinfo {author} {\bibfnamefont {H.}~\bibnamefont {Ishikawa}},
  \bibinfo {author} {\bibfnamefont {S.}~\bibnamefont {Utsunomiya}}, \bibinfo
  {author} {\bibfnamefont {K.}~\bibnamefont {Aihara}}, \ and\ \bibinfo {author}
  {\bibfnamefont {Y.}~\bibnamefont {Yamamoto}},\ }\href@noop {} {\bibfield
  {journal} {\bibinfo  {journal} {Quantum Science and Technology}\ }\textbf
  {\bibinfo {volume} {2}},\ \bibinfo {pages} {044002} (\bibinfo {year}
  {2017})}\BibitemShut {NoStop}%
\bibitem [{\citenamefont {Hamerly}\ \emph {et~al.}(2019)\citenamefont
  {Hamerly}, \citenamefont {Inagaki}, \citenamefont {McMahon}, \citenamefont
  {Venturelli}, \citenamefont {Marandi}, \citenamefont {Onodera}, \citenamefont
  {Ng}, \citenamefont {Langrock}, \citenamefont {Inaba}, \citenamefont {Honjo}
  \emph {et~al.}}]{hamerly2019experimental}%
  \BibitemOpen
  \bibfield  {author} {\bibinfo {author} {\bibfnamefont {R.}~\bibnamefont
  {Hamerly}}, \bibinfo {author} {\bibfnamefont {T.}~\bibnamefont {Inagaki}},
  \bibinfo {author} {\bibfnamefont {P.~L.}\ \bibnamefont {McMahon}}, \bibinfo
  {author} {\bibfnamefont {D.}~\bibnamefont {Venturelli}}, \bibinfo {author}
  {\bibfnamefont {A.}~\bibnamefont {Marandi}}, \bibinfo {author} {\bibfnamefont
  {T.}~\bibnamefont {Onodera}}, \bibinfo {author} {\bibfnamefont
  {E.}~\bibnamefont {Ng}}, \bibinfo {author} {\bibfnamefont {C.}~\bibnamefont
  {Langrock}}, \bibinfo {author} {\bibfnamefont {K.}~\bibnamefont {Inaba}},
  \bibinfo {author} {\bibfnamefont {T.}~\bibnamefont {Honjo}},  \emph
  {et~al.},\ }\href@noop {} {\bibfield  {journal} {\bibinfo  {journal} {Science
  advances}\ }\textbf {\bibinfo {volume} {5}},\ \bibinfo {pages} {eaau0823}
  (\bibinfo {year} {2019})}\BibitemShut {NoStop}%
\bibitem [{\citenamefont {Harrigan}\ \emph {et~al.}(2021)\citenamefont
  {Harrigan}, \citenamefont {Sung}, \citenamefont {Neeley}, \citenamefont
  {Satzinger}, \citenamefont {Arute}, \citenamefont {Arya}, \citenamefont
  {Atalaya}, \citenamefont {Bardin}, \citenamefont {Barends}, \citenamefont
  {Boixo} \emph {et~al.}}]{harrigan2021quantum}%
  \BibitemOpen
  \bibfield  {author} {\bibinfo {author} {\bibfnamefont {M.~P.}\ \bibnamefont
  {Harrigan}}, \bibinfo {author} {\bibfnamefont {K.~J.}\ \bibnamefont {Sung}},
  \bibinfo {author} {\bibfnamefont {M.}~\bibnamefont {Neeley}}, \bibinfo
  {author} {\bibfnamefont {K.~J.}\ \bibnamefont {Satzinger}}, \bibinfo {author}
  {\bibfnamefont {F.}~\bibnamefont {Arute}}, \bibinfo {author} {\bibfnamefont
  {K.}~\bibnamefont {Arya}}, \bibinfo {author} {\bibfnamefont {J.}~\bibnamefont
  {Atalaya}}, \bibinfo {author} {\bibfnamefont {J.~C.}\ \bibnamefont {Bardin}},
  \bibinfo {author} {\bibfnamefont {R.}~\bibnamefont {Barends}}, \bibinfo
  {author} {\bibfnamefont {S.}~\bibnamefont {Boixo}},  \emph {et~al.},\
  }\href@noop {} {\bibfield  {journal} {\bibinfo  {journal} {Nature Physics}\
  }\textbf {\bibinfo {volume} {17}},\ \bibinfo {pages} {332} (\bibinfo {year}
  {2021})}\BibitemShut {NoStop}%
\bibitem [{\citenamefont {B{\"o}hm}\ \emph {et~al.}(2019)\citenamefont
  {B{\"o}hm}, \citenamefont {Verschaffelt},\ and\ \citenamefont {Van~der
  Sande}}]{bohm2019poor}%
  \BibitemOpen
  \bibfield  {author} {\bibinfo {author} {\bibfnamefont {F.}~\bibnamefont
  {B{\"o}hm}}, \bibinfo {author} {\bibfnamefont {G.}~\bibnamefont
  {Verschaffelt}}, \ and\ \bibinfo {author} {\bibfnamefont {G.}~\bibnamefont
  {Van~der Sande}},\ }\href@noop {} {\bibfield  {journal} {\bibinfo  {journal}
  {Nature communications}\ }\textbf {\bibinfo {volume} {10}},\ \bibinfo {pages}
  {1} (\bibinfo {year} {2019})}\BibitemShut {NoStop}%
\bibitem [{\citenamefont {Arora}\ \emph {et~al.}(2005)\citenamefont {Arora},
  \citenamefont {Berger}, \citenamefont {Elad}, \citenamefont {Kindler},\ and\
  \citenamefont {Safra}}]{arora2005non}%
  \BibitemOpen
  \bibfield  {author} {\bibinfo {author} {\bibfnamefont {S.}~\bibnamefont
  {Arora}}, \bibinfo {author} {\bibfnamefont {E.}~\bibnamefont {Berger}},
  \bibinfo {author} {\bibfnamefont {H.}~\bibnamefont {Elad}}, \bibinfo {author}
  {\bibfnamefont {G.}~\bibnamefont {Kindler}}, \ and\ \bibinfo {author}
  {\bibfnamefont {M.}~\bibnamefont {Safra}},\ }in\ \href@noop {} {\emph
  {\bibinfo {booktitle} {46th Annual IEEE FOCS Symposium}}}\ (\bibinfo
  {organization} {IEEE},\ \bibinfo {year} {2005})\ pp.\ \bibinfo {pages}
  {206--215}\BibitemShut {NoStop}%
\bibitem [{\citenamefont {{Gurobi Optimization, LLC}}(2023)}]{gurobi}%
  \BibitemOpen
  \bibfield  {author} {\bibinfo {author} {\bibnamefont {{Gurobi Optimization,
  LLC}}},\ }\href {https://www.gurobi.com} {\enquote {\bibinfo {title} {{Gurobi
  Optimizer Reference Manual}},}\ } (\bibinfo {year} {2023})\BibitemShut
  {NoStop}%
\end{thebibliography}%

\end{document}